\documentclass[12pt,preprint]{aastex}
\usepackage{float}
\shorttitle{ACE observations of the anisotropic solar wind}
\shortauthors{R. M. Nicol, S. C. Chapman and R. O. Dendy}
\begin{document}
\bibliographystyle{plainnat}
\title{QUANTIFYING THE ANISOTROPY AND SOLAR CYCLE DEPENDENCE OF ``$1/f$'' SOLAR WIND FLUCTUATIONS OBSERVED BY ACE}
\author{R. M. N\fontshape{sc}\selectfont{icol} \altaffilmark{1}, S. C. C\fontshape{sc}\selectfont{hapman} \altaffilmark{1} and R. O. D\fontshape{sc}\selectfont{endy} \altaffilmark{1} \altaffilmark{2}}
\altaffiltext{1}{Centre for Fusion, Space and Astrophysics, Department of Physics, University of Warwick, Coventry, CV4 7AL, UK}
\email{R.M.Nicol@warwick.ac.uk}
\altaffiltext{2}{UKAEA Culham Division, Culham Science Centre, Abingdon, Oxfordshire, OX14 3DB, UK}

\maketitle

\begin{abstract}
The power spectrum of the evolving solar wind shows evidence of a spectral break between an inertial range of turbulent fluctuations at higher frequencies and a ``$1/f$'' like region at lower frequencies. In the ecliptic plane at $\sim 1$ AU, this break occurs approximately at timescales of a few hours, and is observed in the power spectra of components of velocity and magnetic field. The ``$1/f$'' energy range is of more direct coronal origin than the inertial range, and carries signatures of the complex magnetic field structure of the solar corona, and of footpoint stirring in the solar photosphere. To quantify the scaling properties we use generic statistical methods such as generalised structure functions and PDFs, focusing on solar cycle dependence and on anisotropy with respect to the background magnetic field. We present structure function analysis of magnetic and velocity field fluctuations, using a novel technique to decompose the fluctuations into directions parallel and perpendicular to the mean local background magnetic field. Whilst the magnetic field is close to ``$1/f$'', we show that the velocity field is ``$1/f^{\alpha}$'' with $\alpha\neq1$. For the velocity, the value of $\alpha$ varies between parallel and perpendicular fluctuations and with the solar cycle. There is also variation in $\alpha$ with solar wind speed. We have examined the PDFs in the fast, quiet solar wind and intriguingly, whilst parallel and perpendicular are distinct, both the $\mbox{\boldmath$B$}$ field and velocity show the same PDF of their perpendicular flucutations, which is close to gamma or inverse Gumbel. These results point to distinct physical processes in the corona, and to their mapping out into the solar wind. The scaling exponents obtained constrain the models for these processes.
\end{abstract}
\section{Introduction}
The solar corona expands non-uniformly into space as a supersonic plasma outflow known as the solar wind \citep{parker_58}. The solar wind carries signatures of coronal dynamics as well as locally generated turbulent phenomena, which span a broad range of scales.\\
\textit{In situ} spacecraft observations of fluctuations in solar wind parameters such as velocity and magnetic field (for example, \citet{ruzmaikin_93} in the ecliptic plane and \citet{horbury_95} in polar flows) typically reveal an inertial range (IR) of turbulence with a ``$5/3$'' inverse power-law scaling at high frequencies and a flatter ``$1/f$''-like scaling range at lower frequencies \citep{matthaeus_86}. The breakpoint between these two ranges is seen to evolve radially \citep{bavassano_82, horbury_96a} with the inertial range extending to lower frequencies with increasing radial distance, suggesting a turbulent energy cascade in the solar wind. The solar wind also has a background magnetic field and is therefore a highly anisotropic plasma environment \citep{shebalin_83,oughton_94}. The strength of this background field relative to the amplitude of the fluctuations determines whether the turbulence is ``strong'', i.e the amplitude of fluctuations is comparable to that of the background magnetic field \citep{sridhar_94,goldreich_95} or ``weak'', i.e the background magnetic field is dominant \citep{ng_97,galtier_00}. This inertial range has been extensively studied using timeseries analysis techniques including power spectra \citep{marsch_90a}, probability density functions (PDFs) \citep{marsch_97,padhye_01,bruno_04} and generalised structure funtions (GSFs) \citep[e.g.][]{horbury_97,hnat_05b,sorriso_valvo_07,chapman_07,nicol_08}.\\
In this paper we focus on the low frequency ``$1/f^{\alpha}$'' range, where the observed $\alpha\sim1$ for magnetic field fluctuations, which is ubiquitous in the solar wind and seen at all latitudes and radial distances. It is dominated by signatures of coronal origin \citep[see][]{matthaeus_86}, unlike the turbulence seen at higher frequency, which is locally evolving. The location of the spectral breakpoint point between the inertial and ``$1/f$'' ranges depends on latitude and radial distance, but it is always possible to see ``$1/f$'' scaling at low frequencies. The power spectral density (PSD) of the ``$1/f$'' range in the interplanetary magnetic field has been extensively studied by \citet{matthaeus_86} and at $1$ AU in the magnitude of the solar wind bulk velocity $\mbox{\boldmath$v$}$ and magnetic field $\mbox{\boldmath$B$}$ by \citet{burlaga_02}. There is also an extensive body of work on the Gaussian and non-Gaussian properties of PDFs of fluctuations in solar wind parameters at these very large scales \citep{marsch_97,burlaga_02,sorriso_valvo_04,bavassano_05} and over a wide range of heliospheric radii. \citet{burlaga_02} used large scale velocity fluctuations at $1$ AU on timescales of one hour to a year to quantify the standard deviation, kurtosis and skewness of PDFs over these scales. Studies of the ``$1/f$'' range in the solar wind thus provide a unique perspective on the physics of coronal processes over the solar cycle. For the first time we consider components of $\mbox{\boldmath$v$}$ and $\mbox{\boldmath$B$}$ defined relative to the local magnetic field, and we systematically distinguish between intervals of fast and slow solar wind at solar maximum and minimum. Here we will focus on the anisotropy of the fluctuations by using a novel decomposition technique, and will take advantage of the long timeseries available from the Advanced Composition Explorer ($ACE$) spacecraft to compare not only fast and slow solar wind streams but also periods of minimum and maximum solar activity.\\
In the inertial range, vector quantities such as \textit{in situ} velocity and magnetic field can be resolved for components both parallel and perpendicular with respect to the background magnetic field $\mbox{\boldmath$B$}$. The duration of the timescale over which the background field is computed is important and both large scale $\mbox{\boldmath$B$}$ \citep{matthaeus_90} and average local $\mbox{\boldmath$B$}$ as a function of the scale of the fluctuations \citep{chapman_07, horbury_08} have been considered in the context of inertial range turbulence. In terms of quantifying scaling, these approaches are generic and the focus of the present paper is to incorporate these ideas in statistical studies of the ``$1/f$'' range, since we anticipate that coronal processes and the transport or propagation of coronal structures will depend on orientation with respect to the background magnetic field. The observed scaling would also be anticipated to depend quantitatively on solar cycle and to differ between fast ($\sim 750$ km/s) or slow ($\sim 350$km/s) solar wind streams. High speed flows originate in coronal holes \citep{krieger_73}, whereas low speed flows arise from dense coronal streamers \citep{gosling_81}, while solar rotation causes high and low speed flows to interact at low latitudes. We will perform generalised structure function analysis (GSF) \citep{sornette_04} on datasets spanning these intervals in order to quantify the scaling properties of the magnetic and velocity field fluctuations both parallel and perpendicular to the background magnetic field $\mbox{\boldmath$B$}$. \\
The location of the spectral breakpoint between the inertial and ``$1/f$'' ranges differs in fast and slow streams \citep{horbury_05,bruno_05}, presumably because at a given heliocentric distance the turbulence in the slow solar wind has had more time to develop than in the fast solar wind. Furthermore, the crossover between IR and ``$1/f$'' is much clearer in fast than in slow solar wind. Here, we will see that projecting velocity and magnetic field parallel and perpendicular to $\mbox{\boldmath$B$}$ provides a clear indicator of where this crossover occurs. We compare the position of this breakpoint in fast and slow solar wind streams and at periods of maximum and minimum solar activity. We first see that the PSDs of the vector components of the velocity $\mbox{\boldmath$v$}$ and magnetic field $\mbox{\boldmath$B$}$ suggest anisotropy in the ``$1/f$'' range. We then decompose $\mbox{\boldmath$v$}$ and $\mbox{\boldmath$B$}$ into parallel and perpendicular fluctuations with respect to the local background magnetic field $\mbox{\boldmath$B$}$. For the simple case of quiet fast solar wind, we compare the PDFs of the fluctuations to see which components may or may not share the same underlying generating process. For completeness, we also consider the PDF for the density fluctuations $\delta\rho$. We compare the GSFs for fast and slow solar wind at solar maximum and minimum. Finally, using the GSFs, we obtain values for the scaling exponents in the ``$1/f$'' range and find that these are clearly distinct for $\delta v_{\parallel,\perp}$ and $\delta b_{\parallel,\perp}$.\\
\section{The datasets}
\begin{figure}[H]
\figurenum{1}
\epsscale{0.4}
\plotone{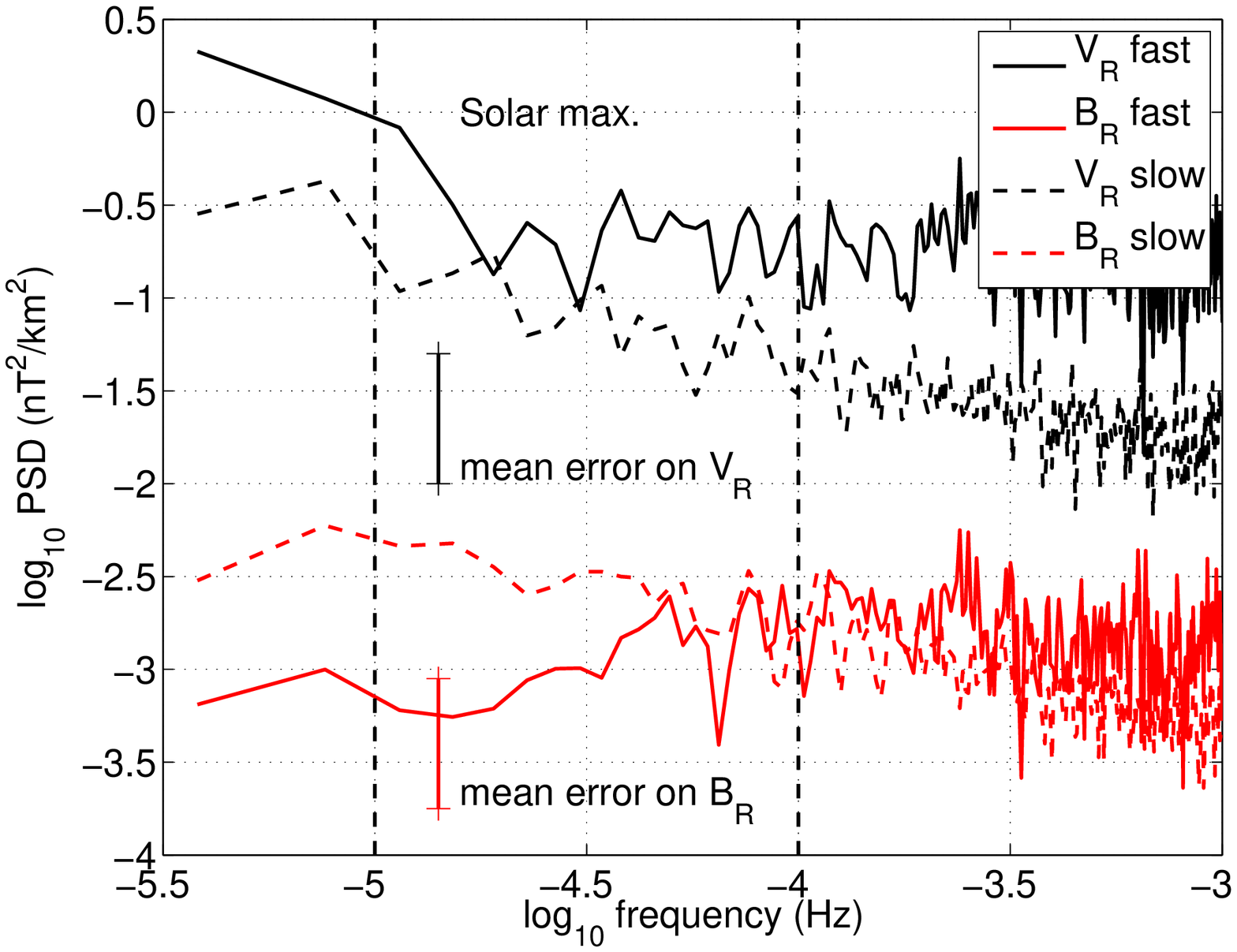}
\plotone{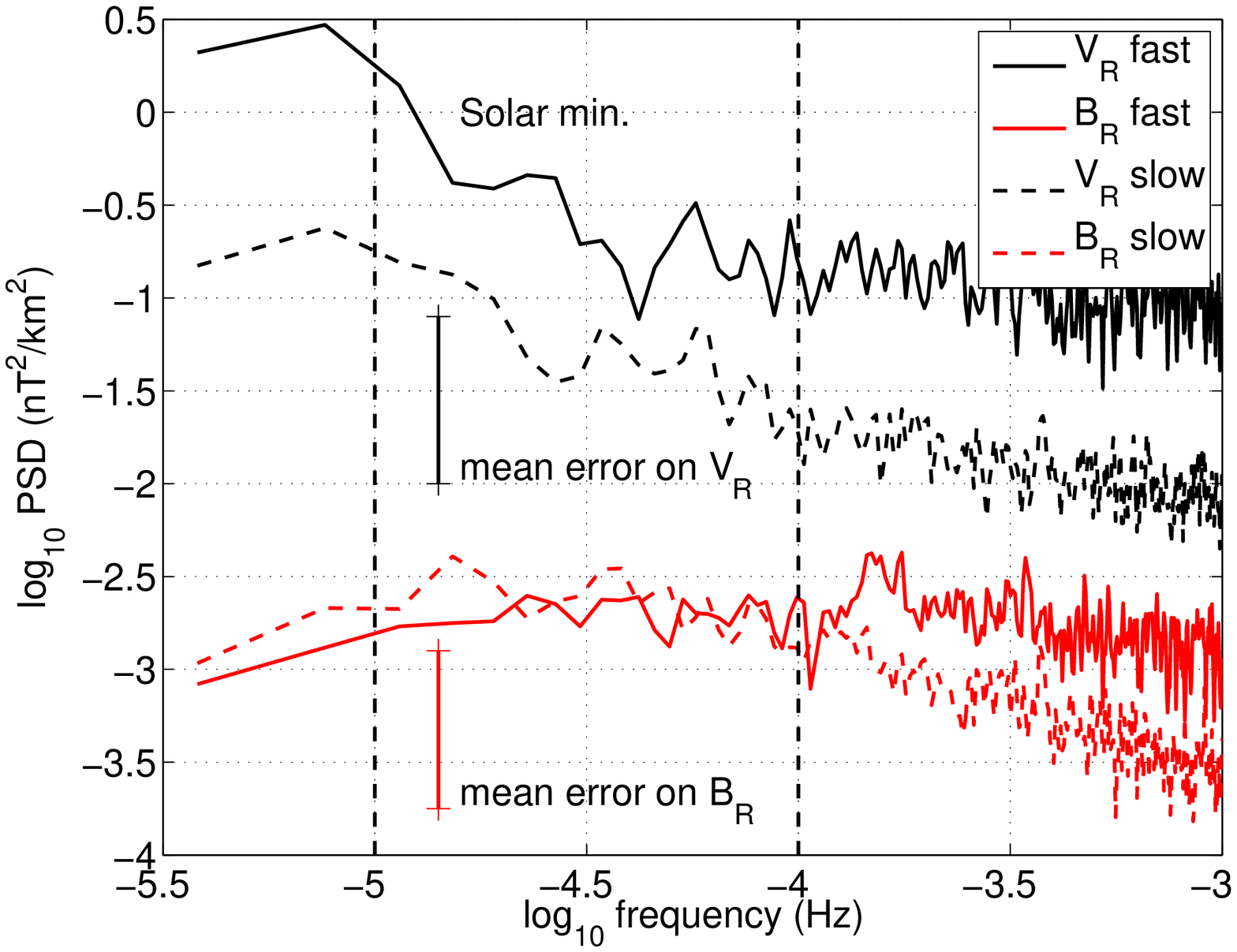}
\plotone{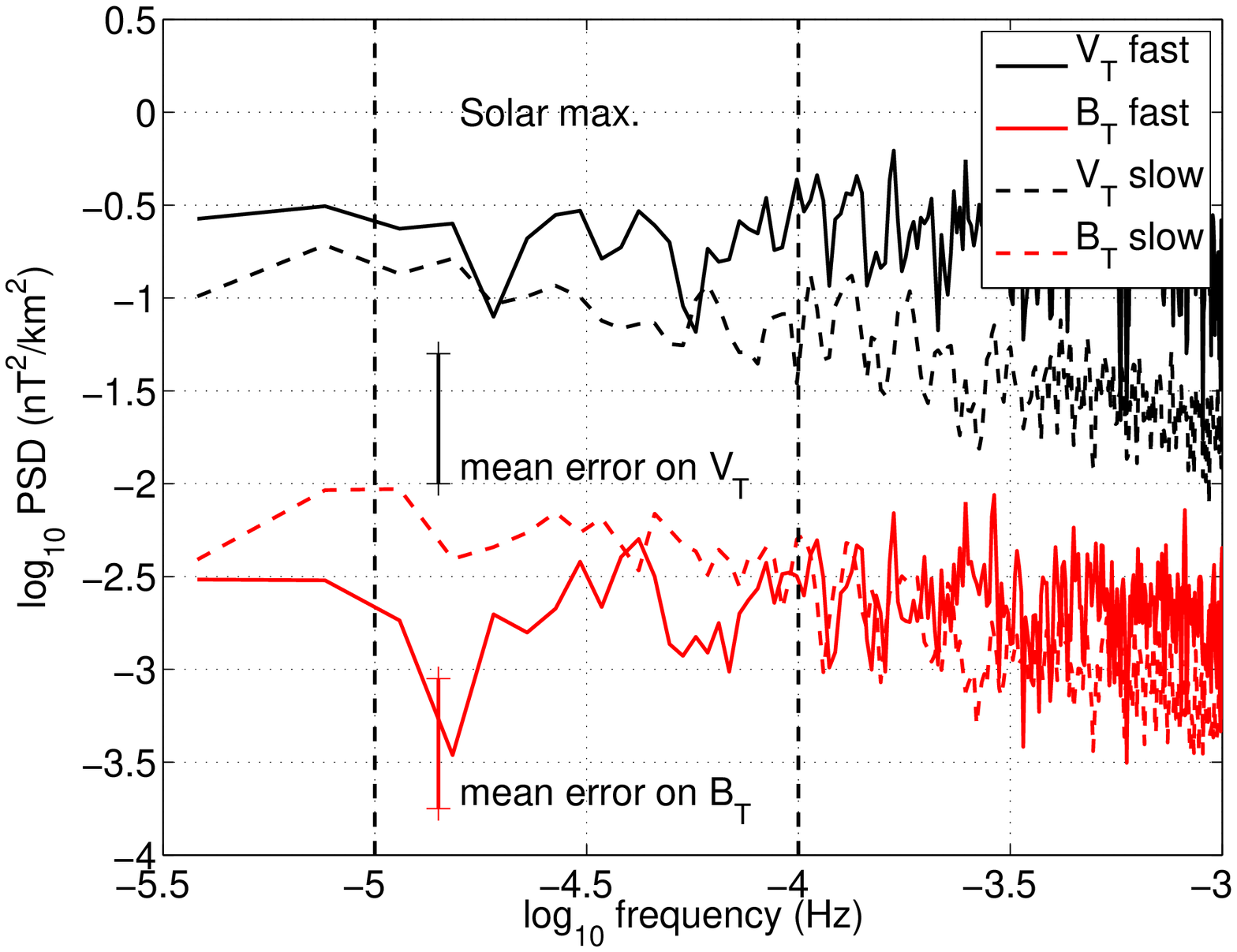}
\plotone{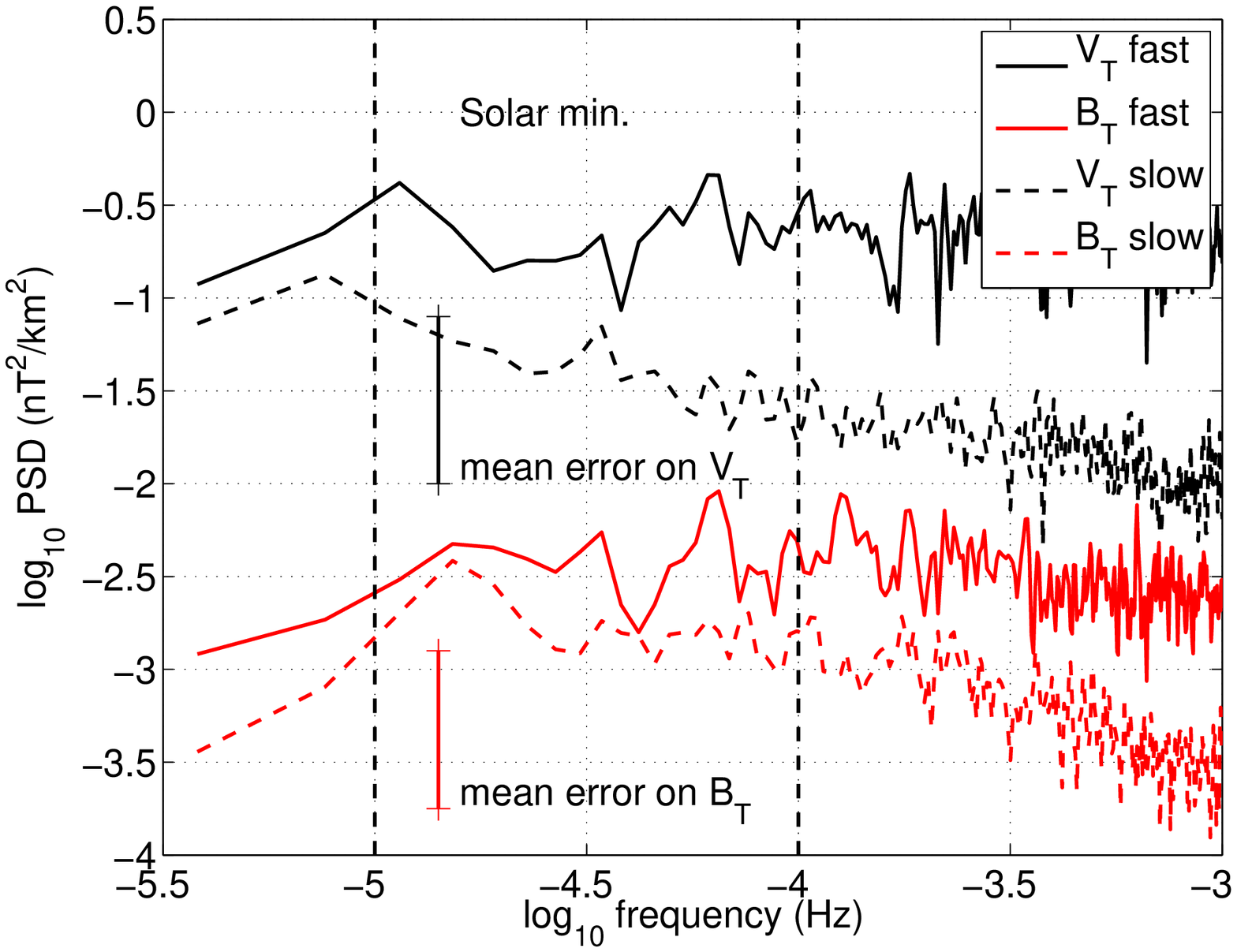}
\plotone{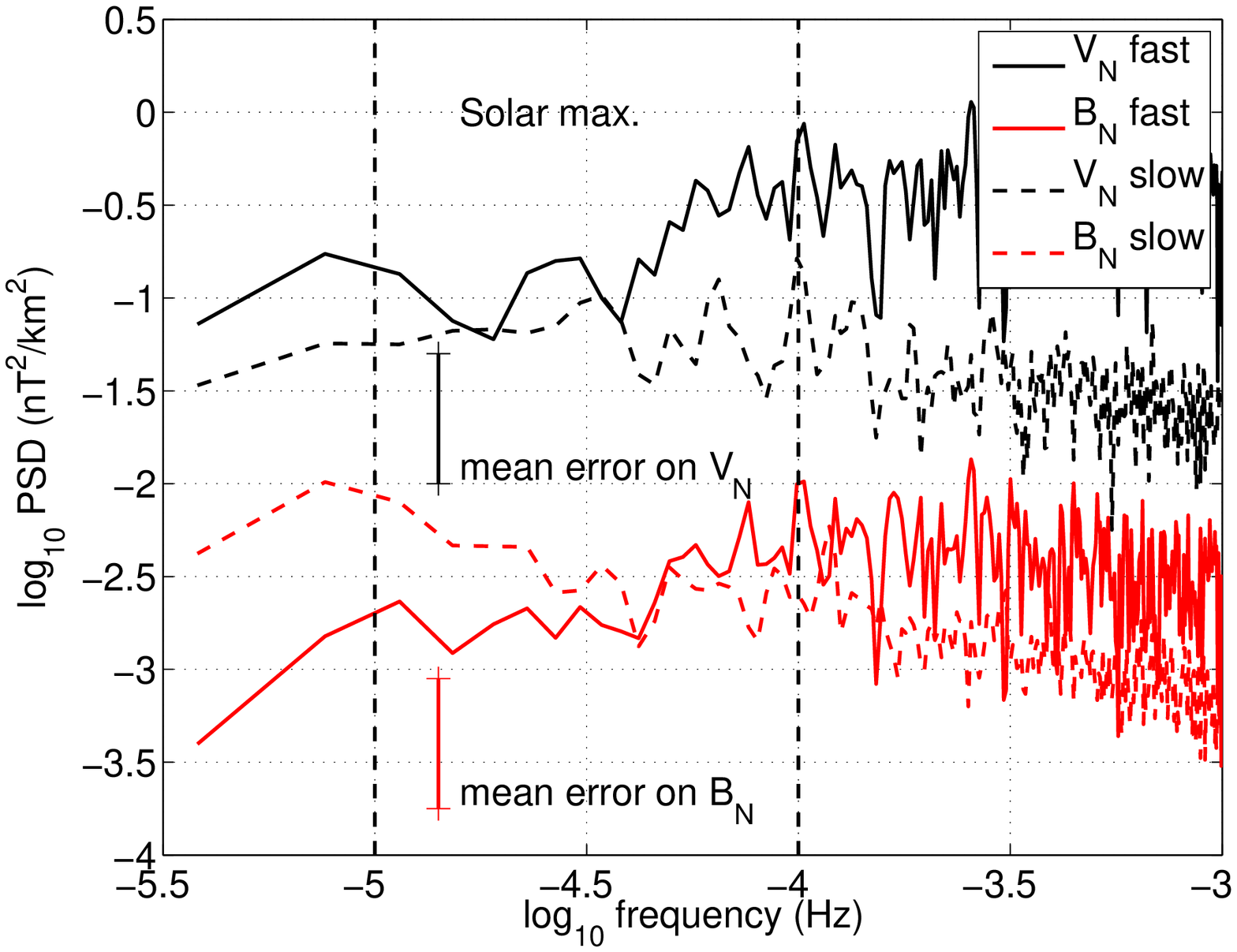}
\plotone{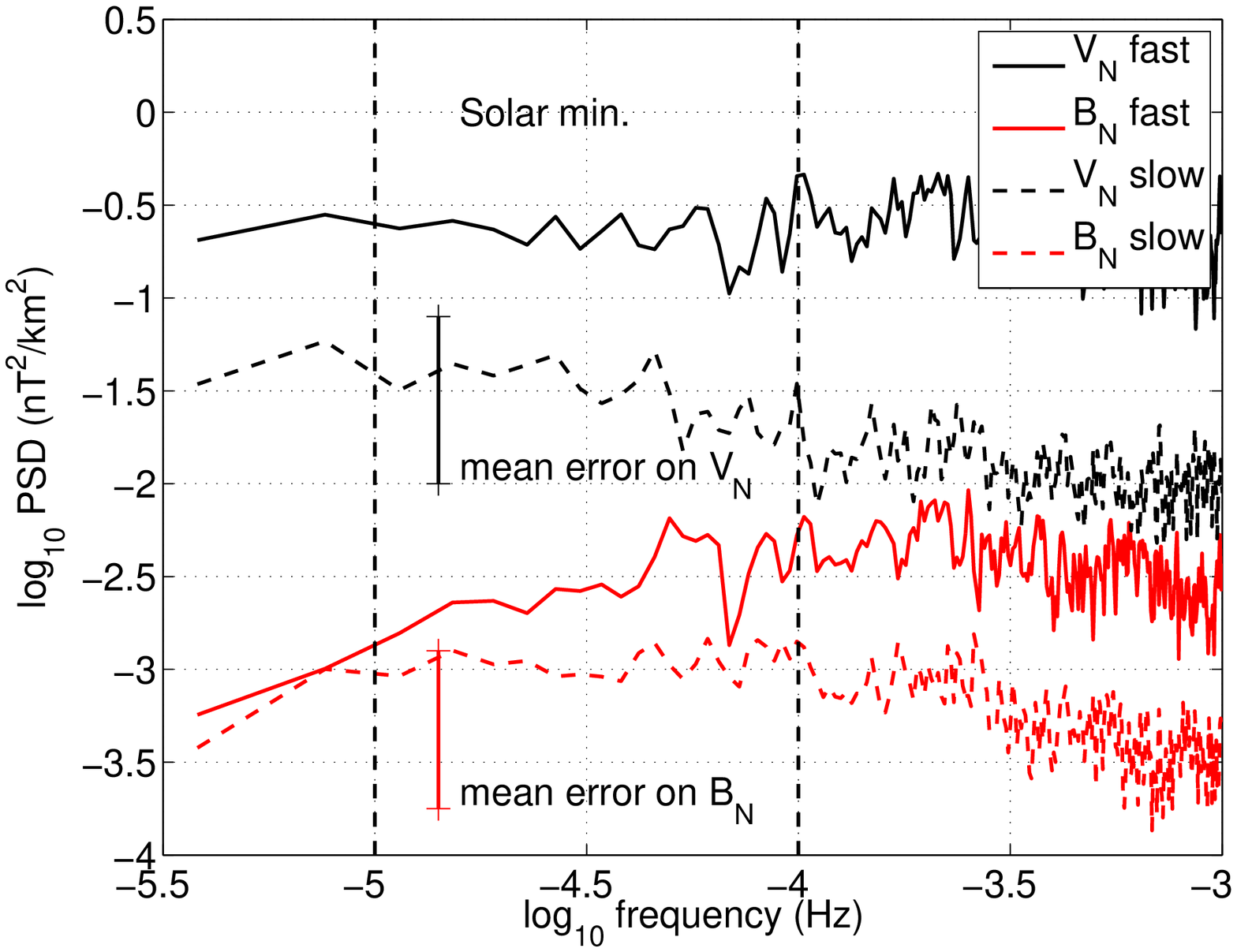}
\caption{Compensated power spectral density $F(f)/f^{\alpha},\;\alpha=-1$ for velocity and magnetic field fluctuation components in the $RTN$ coordinate system for the frequency range $10^{-5.5}-10^{-3}$Hz. Results for the fast (continous line) and slow (dashed line) are displayed separately. The dotted vertical lines delimit the frequency range $10^{-5}-10^{-4}$Hz; this is expected to lie within the ``$1/f$'' range, with the breakpoint between the inertial and ``$1/f$'' ranges $\sim10^{-4}$Hz \citep{marsch_90b,horbury_96a}. The three panels on the left-hand side are for solar maximum, while the right-hand side is solar minimum. The errors are found by considering one standard deviation of the datasets over which the averages are taken.}
\label{Fig.1}
\end{figure}

The advanced composition explorer (ACE) spacecraft \citep{stone_98} orbits the Lagrangian point sunwards of the earth ($\sim1$AU). For the present analysis we study plasma parameters (magnetic field $\mbox{\boldmath$B$}$ and velocity $\mbox{\boldmath$v$}$) averaged over $64$ seconds from the MAG/SWEPAM teams \citep{smith_98,mccomas_98}: for the year $2007$, representative of a period of minimum solar activity; and for the year $2000$, which was a period of maximum solar activity. This provides datasets of $\sim 4.8 \times 10^5$ samples per year. In order to separate fast and slow solar wind behaviour yet still preserve a dataset with sufficient points to perform GSF to explore the ``$1/f$'' dynamic frequency range, we divide the datasets into intervals ($\geq 6000$ points or $4.5$ days) of fast and slow streams, where the cut-off between fast and slow is taken at $450$ km/s \citep[e.g.][]{horbury_05}. These intervals then form one long fast solar wind dataset of $\sim 7.4\times10^{4}$ points, and one long slow solar wind dataset of $\sim 1.4\times10^{5}$ points for the year $2007$ and a fast dataset of $\sim 4.1\times10^{4}$ points and a slow dataset of $\sim 1.1\times10^{5}$ points for the year $2000$. To evaluate spectral properties, we apply Fourier techniques to the original continuous intervals of fast and slow solar wind. When we perform statistical analysis using the probability density functions (PDFs) of fluctuations in section $3$, each dataset is treated as a single ensemble. As we preserve the time indicators for the data, the pairs of datapoints are always drawn from within continuous intervals of fast or slow streams.\\
We first provide an overview of the ``$1/f$'' range of these data intervals by plotting the power spectral density $F(f)$ of the components of $\mbox{\boldmath$v$}$ and $\mbox{\boldmath$B$}$ in the $RTN$ coordinate system, where $R$ is the sun-spacecraft axis, $T$ is the cross product of $R$ with the solar rotation axis, and $N$ is the cross product of $R$ with $T$. Generally, for a signal $x(t)$ of length $N$, the power spectrum $F(f)$ from the fast Fourier transform (FFT) to frequency space is given by
\begin{equation}
F(f)=\frac{1}{N}\mid\sum^{N}_{t=1}x(t)e^{-2i\pi(t-1)(f-1)/N}\mid^{2}
\end{equation}
for a range of frequencies $f=\frac{n}{Nf_s}$ where $n=[0:N/2]$ and $f_s$ is the sampling frequency. We take our original intervals of fast and slow solar wind and truncate (or cut) them such that they all have the same length of $6000$ datapoints. Each interval is then split up into windows of $2^{12}=4096$ points with a $50\%$ overlap on the previous window. A Hamming window is applied to each of these sub-intervals and the FFT is computed. An average is then taken of these sub-interval FFTs to obtain the power spectrum for each interval. The power spectra for all intervals are then averaged to obtain the PSDs for fast and slow solar wind at both solar maximum and minimum. At lower frequencies, the magnetic field power spectrum $F(f)\sim f^{\alpha}$ shows a spectral slope $\alpha\sim-1$. Plotting $F(f)/f^{\alpha},\;\alpha=-1$ should therefore give a horizontal line (on average). These plots are known as compensated power spectra and are shown for the various solar wind conditions in Figure \ref{Fig.1}.
Figure \ref{Fig.1} covers the expected region of transition in the spectral index of $\mbox{\boldmath$v$}$ and $\mbox{\boldmath$B$}$ between the IR and ``$1/f$'' frequency ranges. However it is difficult to tell precisely whether, for example, the PSD behaviour between $10^{-5}$Hz and $10^{-4}$Hz really is ``$1/f^{\alpha}$, $\alpha=1$'', particularly in the slow solar wind. It also evident from Figure \ref{Fig.1} that in some cases in the ``$1/f^{\alpha}$'' range $\alpha$ varies with the solar cycle and with solar wind speed, and that for both $\mbox{\boldmath$v$}$ and $\mbox{\boldmath$B$}$ the $\alpha$ can vary from one component to another, and between $\mbox{\boldmath$v$}$ and $\mbox{\boldmath$B$}$. This implies anisotropy in the fluctuations and distinct scaling between $\mbox{\boldmath$v$}$ and $\mbox{\boldmath$B$}$.\\
From a statistical point of view, let us now characterize this anisotropy by decomposing the velocity (or magnetic) field fluctuations into parallel and perpendicular components relative to the background magnetic field. We adopt the Taylor hypothesis \citep{taylor_38} to relate spatial and temporal scales and fluctuations over a time lag $\tau$ in the velocity (or magnetic field) vector components, defined as $\delta \mbox{\boldmath$v$}(t,\tau)=\mbox{\boldmath$v$} (t+\tau)-\mbox{\boldmath$v$}(t)$. A vector average for the magnetic field direction $\mbox{\boldmath$\hat b$}(t,\tau)=\mbox{\boldmath$\overline B$}/\vert\mbox{\boldmath$\overline B$}\vert$ is formed from a vector sum $\mbox{\boldmath$\overline B$}(t)$ of all the observed vector $\mbox{\boldmath$B$}$ values between $t-\tau/2$ and $t+3\tau/2$. It follows that in computing fluctuations over $\tau$, the background field is averaged over $\tau'=2\tau$, which then defines the minimum (Nyquist) interval necessary to capture wavelike fluctuations \citep{chapman_07}. Using this definition of $\mbox{\boldmath$\hat b$}$, the inner product
\begin{equation}
\delta v_{\parallel}=\delta\mbox{\boldmath$v$}\cdot\mbox{\boldmath$\hat b$}=\delta v_R\hat b_R+\delta v_T\hat b_T+\delta v_N\hat b_N
\end{equation}
vanishes for fluctuations which generate a velocity displacement that is purely perpendicular to the background magnetic field $\mbox{\boldmath$\overline B$}$ as defined. The perpendicular fluctuation amplitude is then obtained from
\begin{equation}
\delta v_{\perp}=\sqrt{\delta\mbox{\boldmath$v$}\cdot\delta\mbox{\boldmath$v$}-\left( \delta\mbox{\boldmath$v$}\cdot\mbox{\boldmath$\hat b$}\right) ^{2}}
\end{equation}
We use these definitions to construct differenced timeseries $\delta v_{\perp}(t, \tau)$, $\delta b_{\perp}(t, \tau)$, $\delta v_{\parallel}(t, \tau)$ and $\delta b_{\parallel}(t, \tau)$ over a range of $\tau$ intervals within the ``$1/f$'' range, that is $\tau$ from a few hours up to a day.\\
\section{PDF analysis}
\begin{figure}[H]
\figurenum{2}
\epsscale{0.40}
\plotone{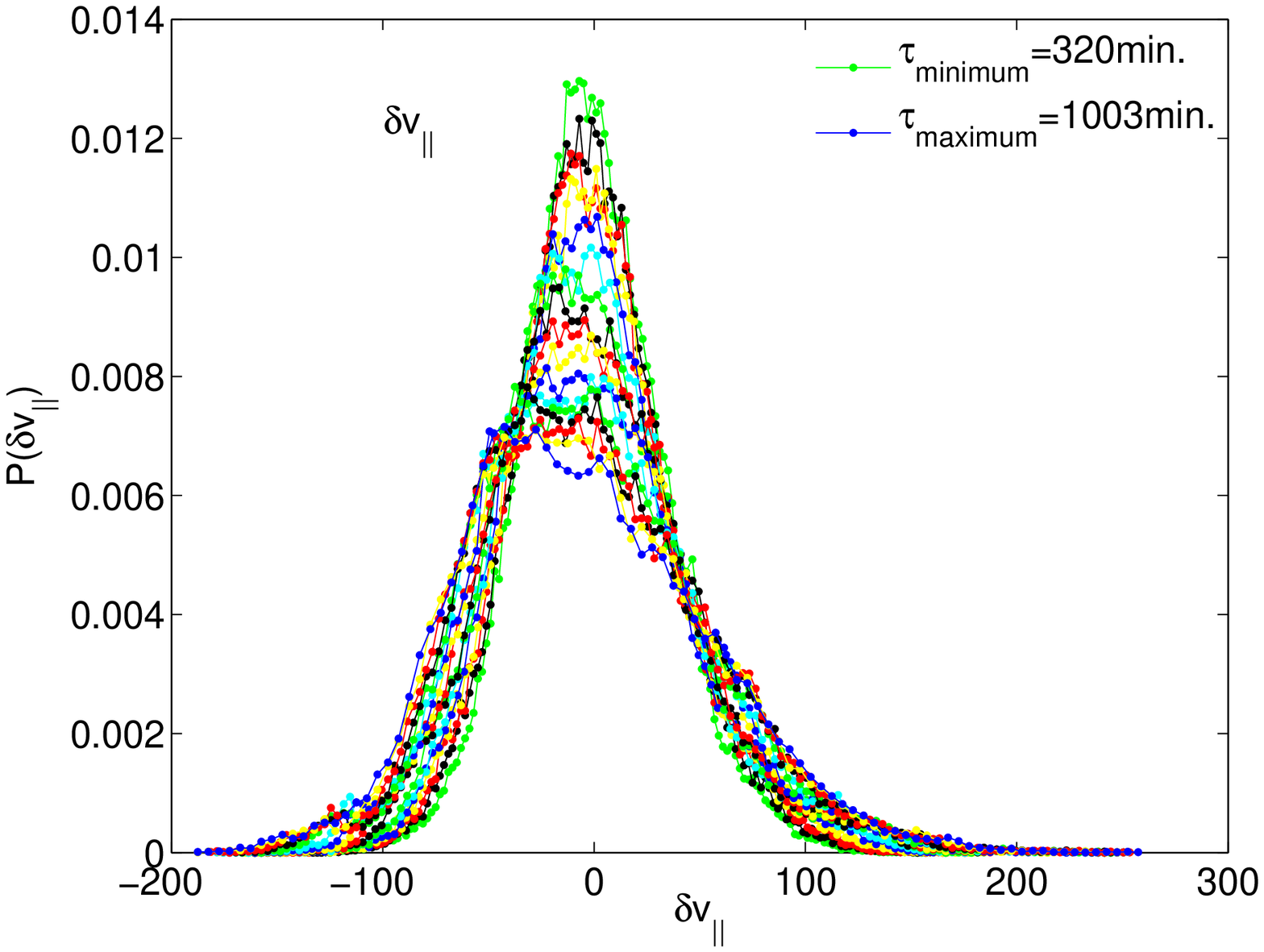}
\plotone{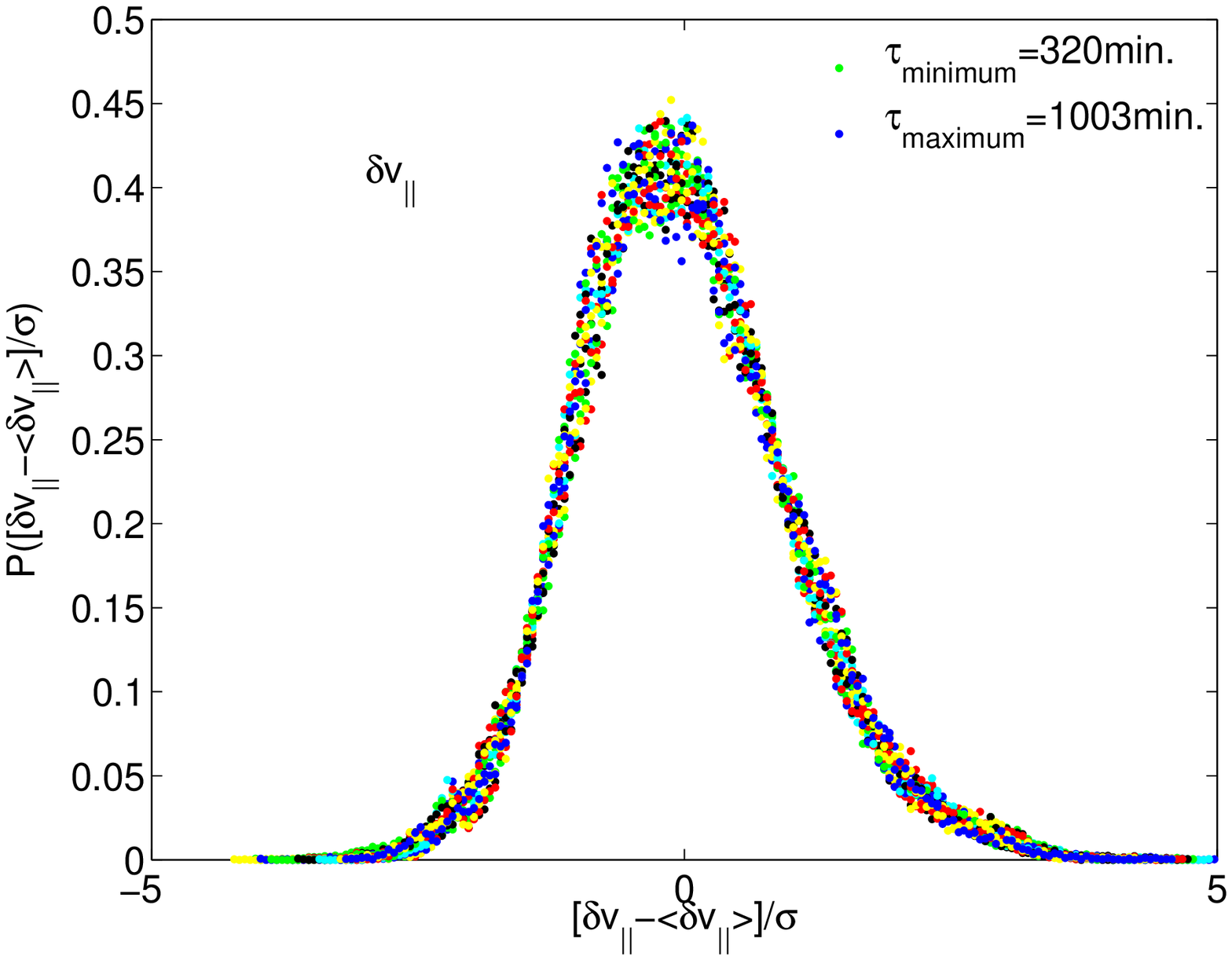}
\plotone{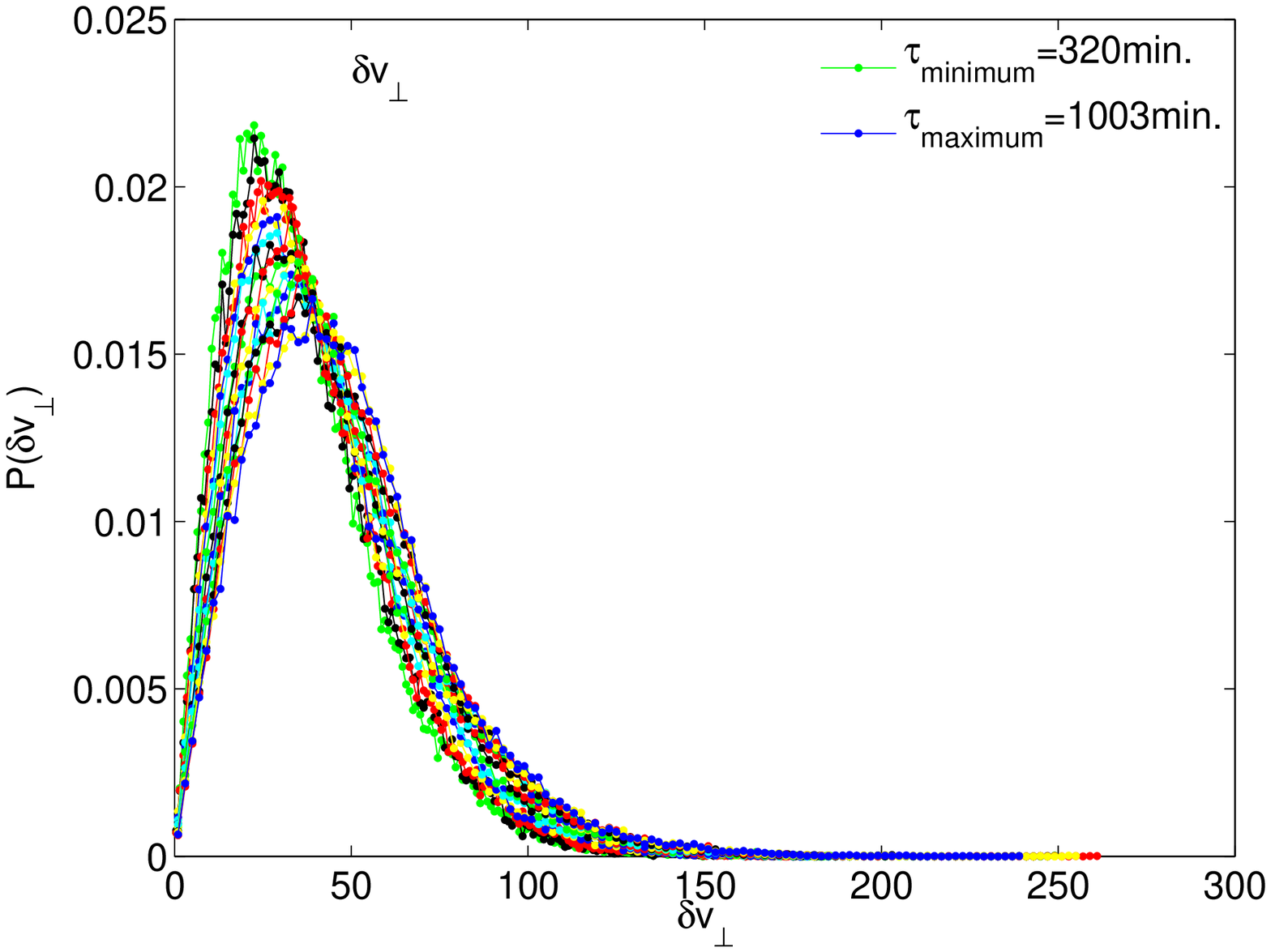}
\plotone{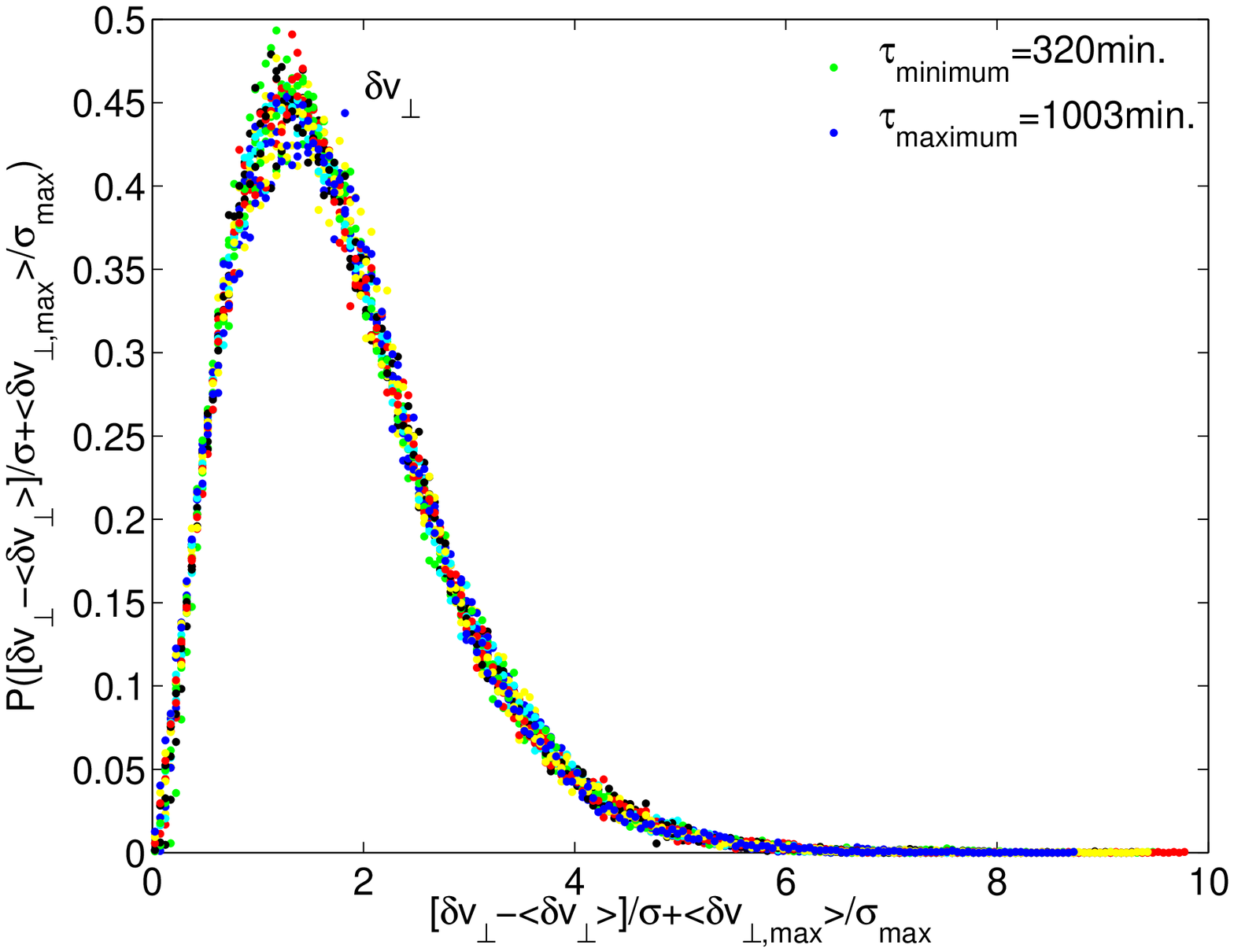}
\caption{Parallel (upper) and perpendicular (lower) velocity fluctuations $\delta v_{\parallel,\perp}$ in the fast solar wind at solar minimum for the ``$1/f$'' range. The left panels show the PDFs of raw fluctuations sampled across intervals $\tau$ between $320$ and $1003$ minutes. The right panels show the same curves normalised using equation \ref{eqn6}.}
\label{Fig.2}
\end{figure}

\begin{figure}[H]
\figurenum{3}
\epsscale{0.40}
\plotone{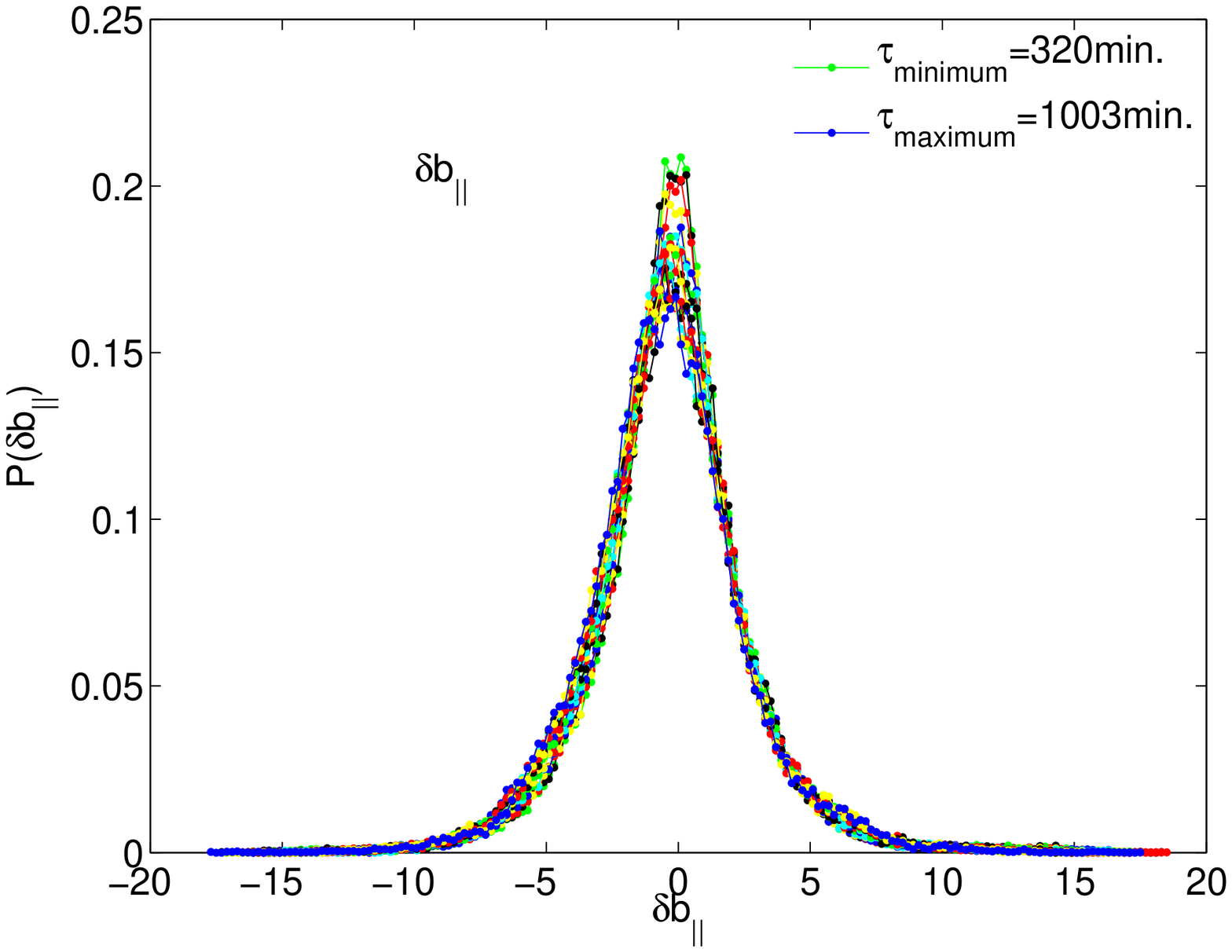}
\plotone{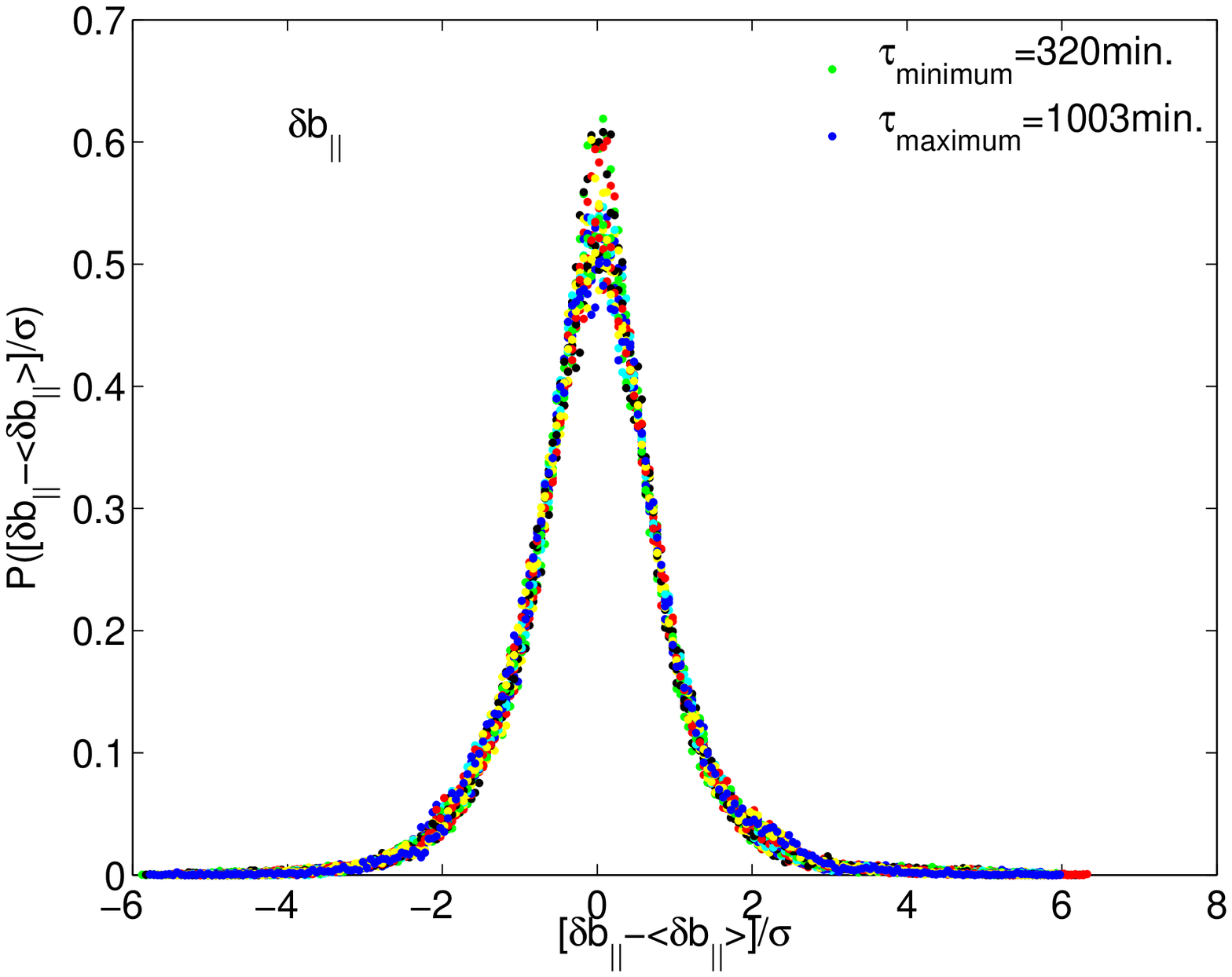}
\plotone{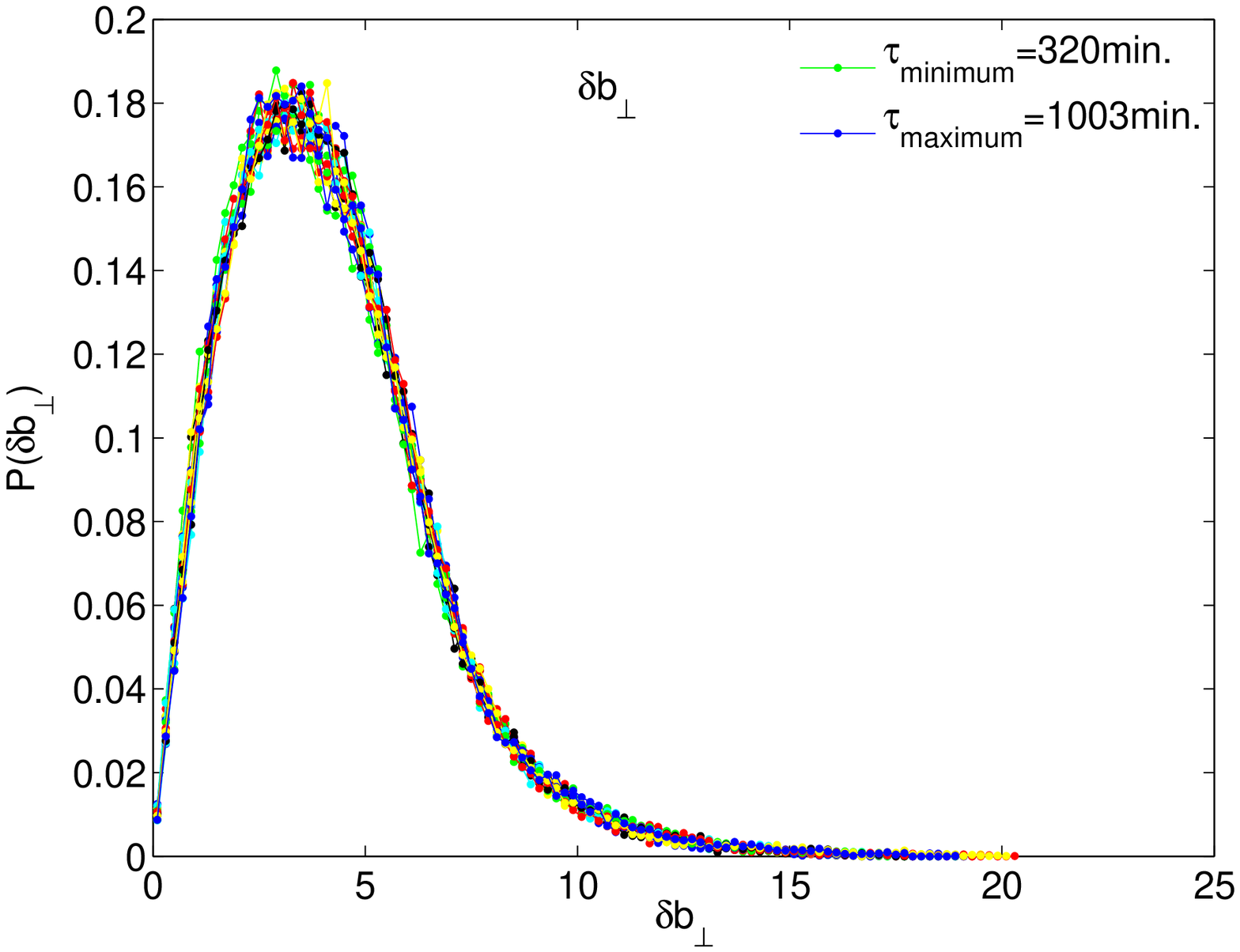}
\plotone{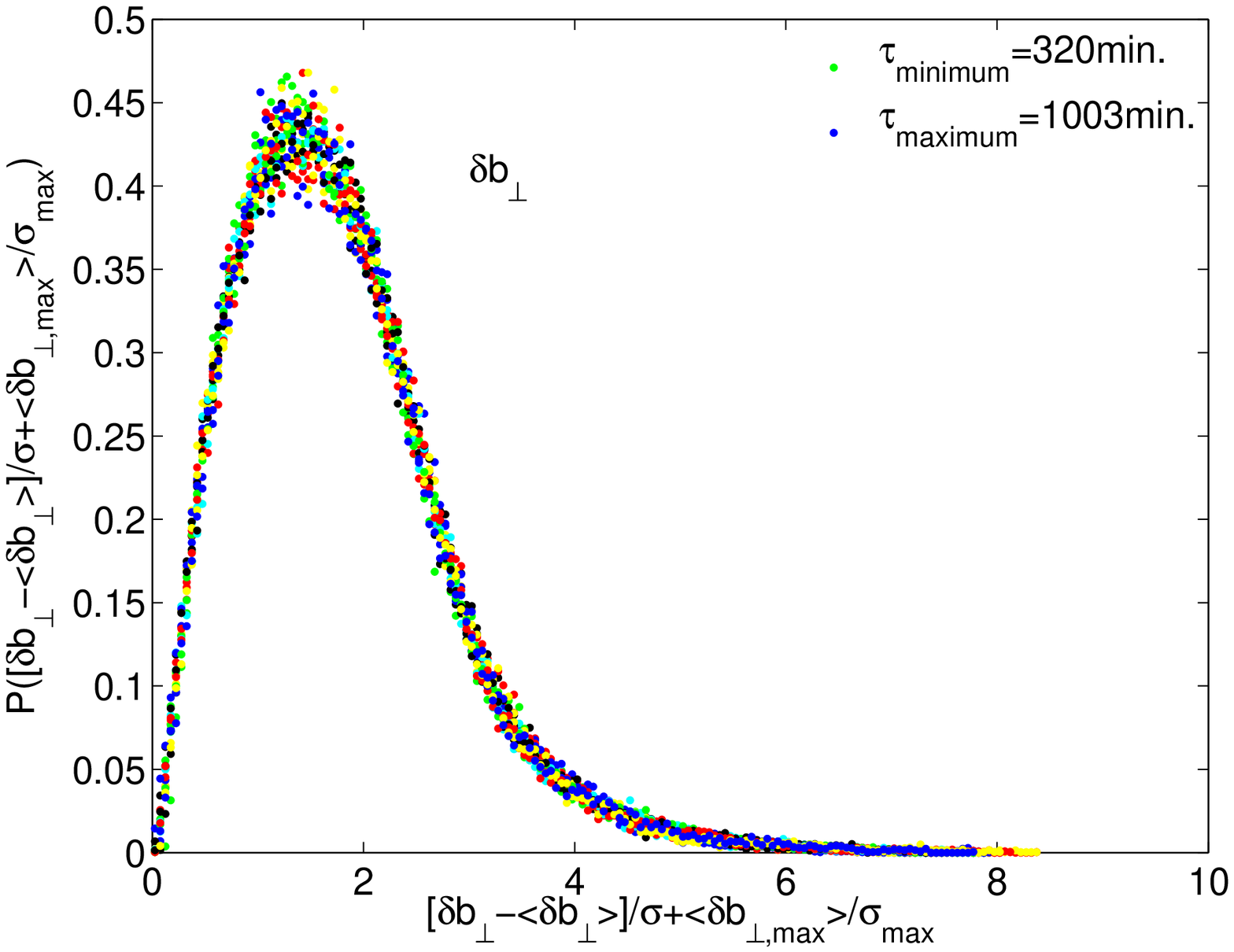}
\caption{Parallel (upper) and perpendicular (lower) magnetic field fluctuations $\delta b_{\parallel,\perp}$ in the fast solar wind at solar minimum for the ``$1/f$'' range. The left panels show the PDFs of raw fluctuations sampled across intervals $\tau$ between $320$ and $1003$ minutes. The right panels show the same curves normalised using equation \ref{eqn6}.}
\label{Fig.3}
\end{figure}

\begin{figure}[H]
\figurenum{4}
\epsscale{0.40}
\plotone{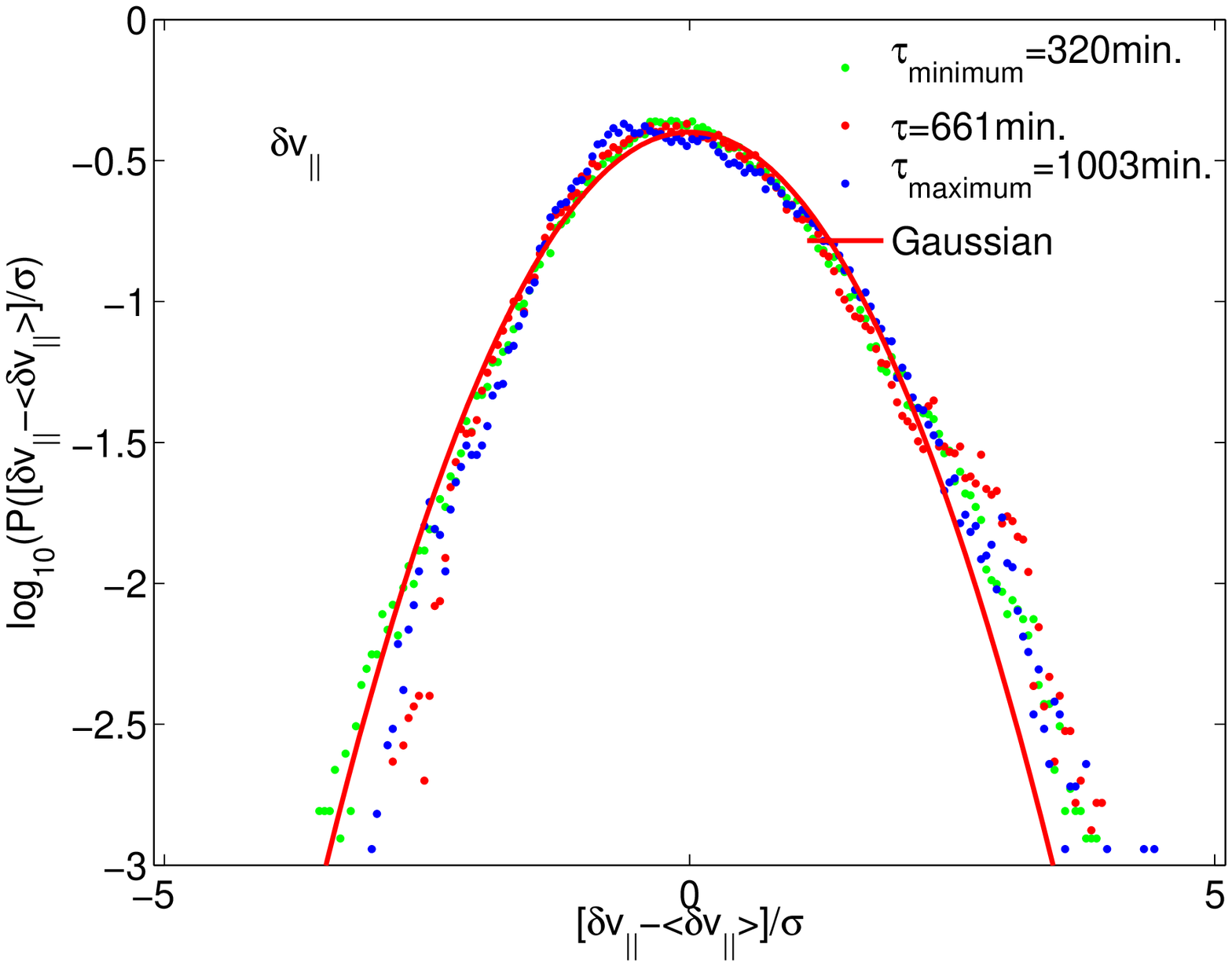}
\plotone{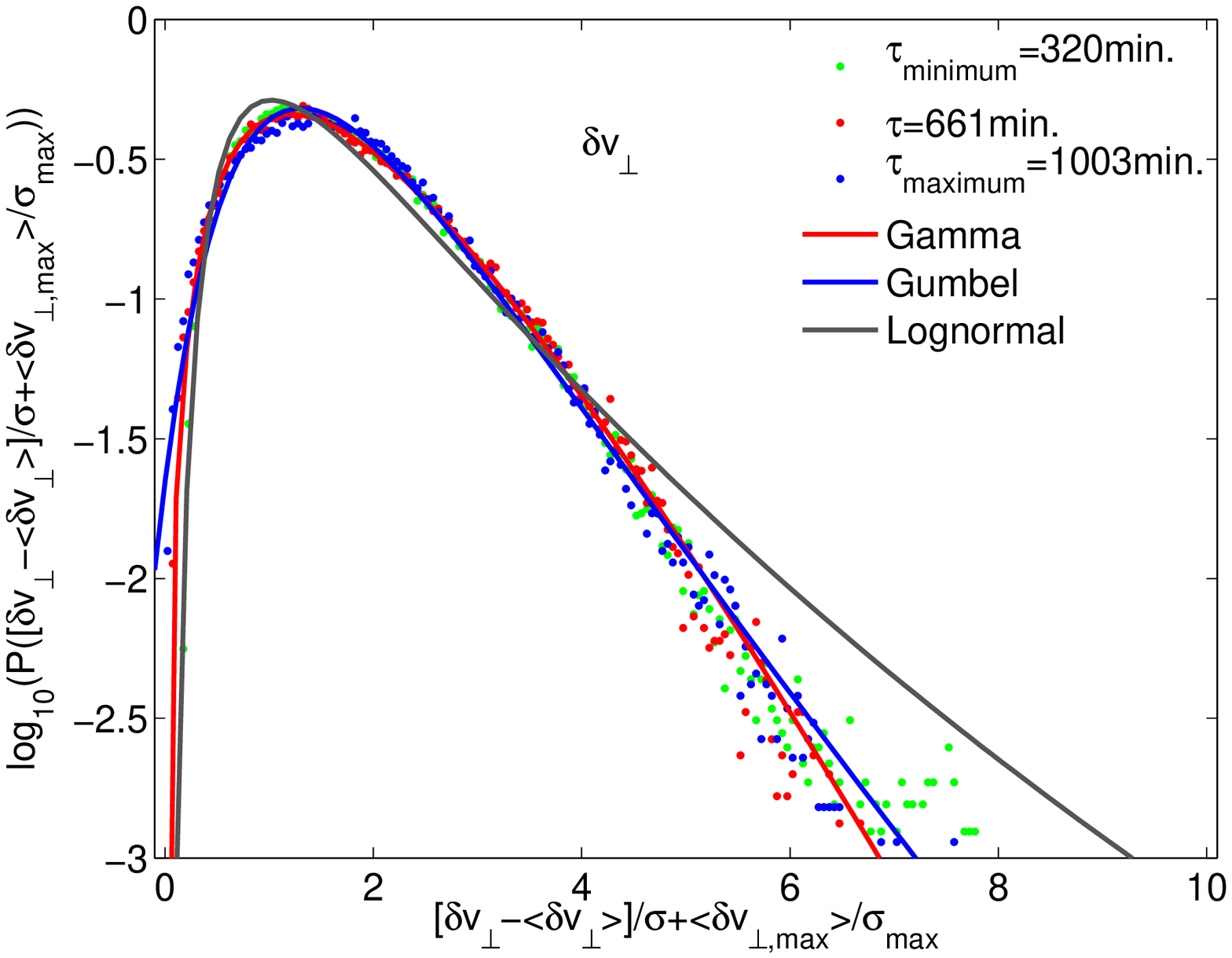}
\plotone{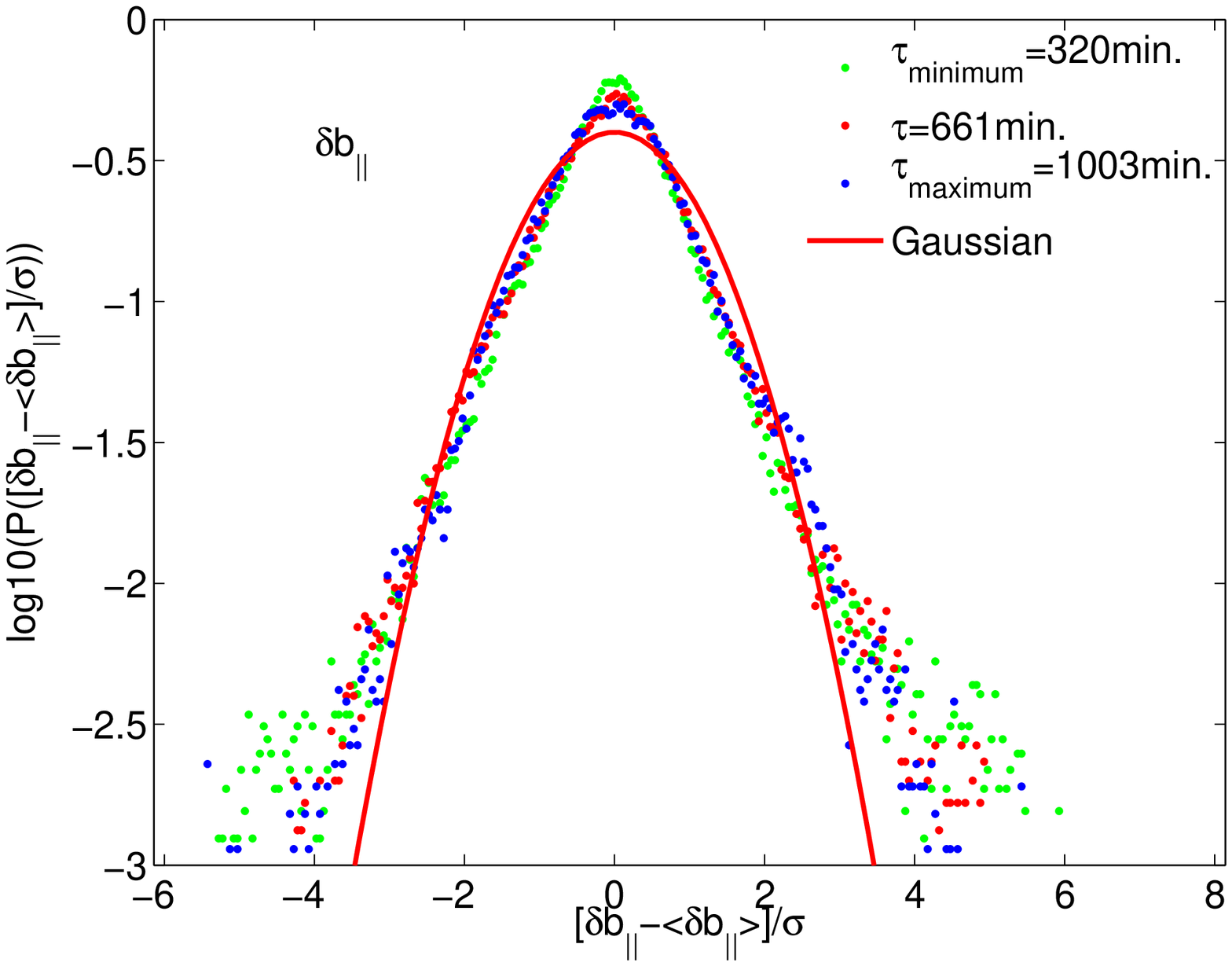}
\plotone{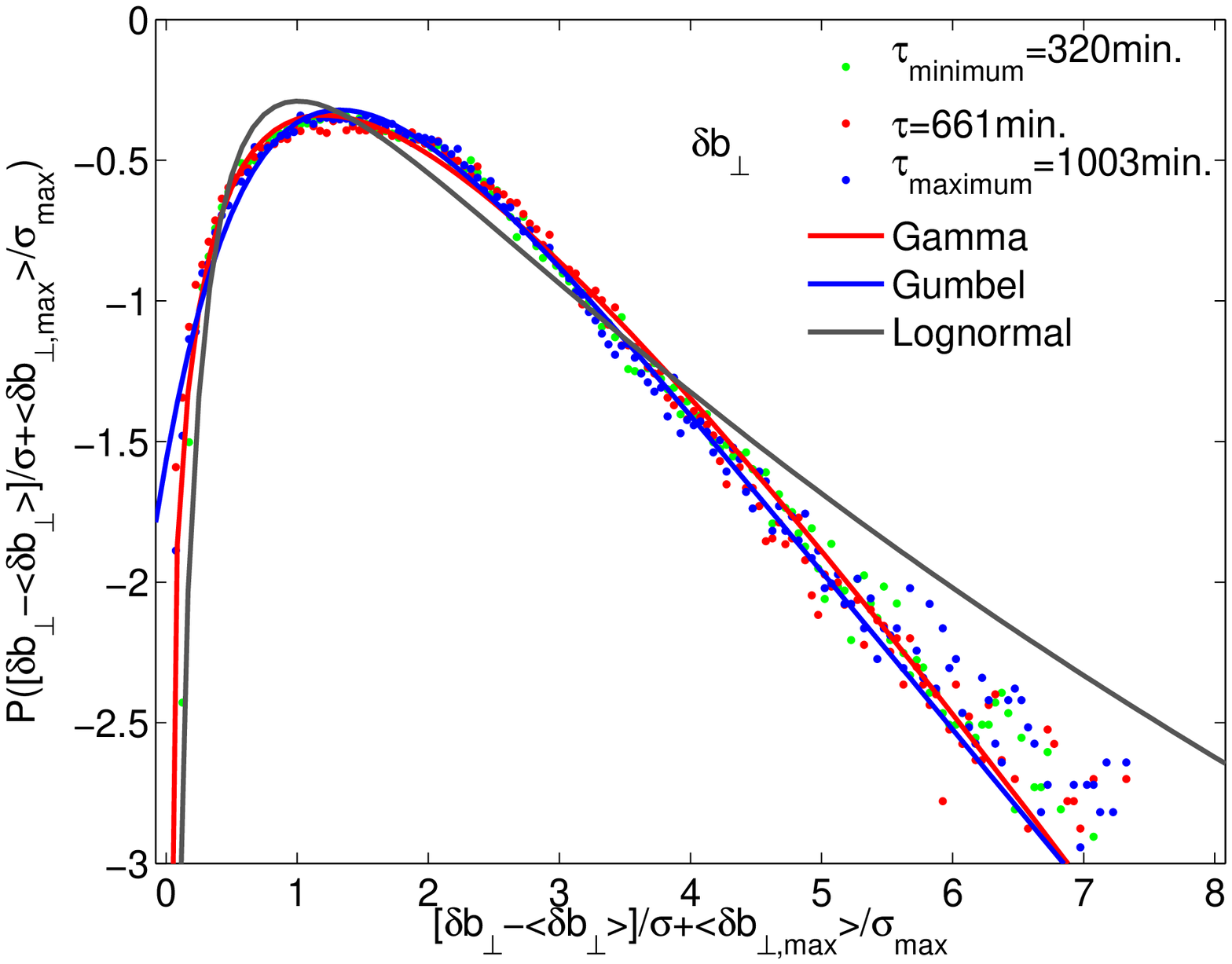}
\caption{Parallel and perpendicular velocity and magnetic field fluctuations $\delta v_{\parallel,\perp}$ and $\delta b_{\parallel,\perp}$ in the fast solar wind at solar minimum for the ``$1/f$'' range. For clarity, plots for only three representative values of $\tau$ are shown for each component, whereas the fitted curves are computed using all the $\tau$ intervals between $320$ and $1003$ minutes. The left panels show a Gaussian fit to the normalised PDF curves for both $\delta v_{\parallel}$ (upper) and $\delta b_{\parallel}$ (lower) using semilog $y$ axes. The right panels show the normalised curves $\delta v_{\perp}$ (upper) and $\delta b_{\perp}$ (lower) fitted with three different distributions gamma (red); Gumbel (blue); and lognormal (grey) on semilog $y$ axis. For the perpendicular components, the renormalisation with $\mu$ means that the PDFs are shifted so that they are centered on zero, however the gamma distribution can only have positive arguments. It is therefore necessary to shift the PDFs by $<\delta v_{\perp,max}>/\sigma_{max}$.}
\label{Fig.4}
\end{figure}

We first examine the PDFs of these fluctuations and explore their possible functional forms. For a self-affine process, knowledge of the functional form of the PDF, and of the Hurst exponent $H$, is sufficient in principle to build a stochastic differential equation model for the process \citep[e.g.][]{sornette_04,chapman_05,kiyani_07}. To compare their functional form, the PDFs can be renormalised using \citep[e.g.][]{greenhough_02}
\begin{equation}
P[(y-<y>)]=\sigma^{-1}P[\sigma^{-1} (y-<y>)]\label{eqn6}
\end{equation}
where $<\cdots>$ denotes the ensemble mean and $\sigma$ is the standard deviation of the distribution. From a statistical point of view, where fluctuations arise from a single physical process, rescaling of PDFs using equation \ref{eqn6} leads to the ``collapse'' of the PDFs for the different $\tau$ onto a single function that characterizes the underlying process \citep[e.g.][]{greenhough_02b,dudson_05,dendy_06,dewhurst_08,hnat_08}. Let us apply this technique to parallel and perpendicular velocity and magnetic field fluctuations in the fast solar wind at solar minimum. Figure \ref{Fig.2} shows that the PDFs for the $\delta v_{\parallel}$ and $\delta v_{\perp}$ components each collapse onto single curves that are distinct from each other. The PDF for $\delta v_{\parallel}$ is asymmetric about $\delta v_{\parallel}=0$, and we have investigated this asymmetry by sorting the fluctuations with respect to the sign of $\delta v_{R}$ into $\delta v_{\parallel}^{+}$ and $\delta v_{\parallel}^{-}$. The resulting GSFs and scaling exponents display the same fractal characteristics as $\delta v_{\parallel}$, implying that $\delta v_{\parallel}^{+}$ and $\delta v_{\parallel}^{-}$ arise from the same physical process.
Figure \ref{Fig.3} shows that the PDFs for $\delta b_{\parallel}$ and $\delta b_{\perp}$ each collapse onto single curves that are distinct from each other. The curve for $\delta b_{\parallel}$ is distinct from that for $\delta v_{\parallel}$ and the PDF has stretched exponential tails, which implies that these fluctuations may originate in multiplicative or fractionating process \citep{frisch_97}. The curves for $\delta b_{\perp}$ and $\delta v_{\perp}$ look remarkably similar and we will explore this later. 
The functional forms of these distributions are investigated in Figure \ref{Fig.4}. A Gaussian distribution \citep{wadsworth_97}
\begin{equation}
\frac{1}{\sigma\sqrt{2\pi}}e^{\frac{-(x-\mu)^2}{2\sigma^2}} 
\end{equation}
approximately fits the normalised PDFs of the $\delta v_{\parallel}$ fluctuations in the ``$1/f$'' range shown in Figure \ref{Fig.2} with the following fitting parameters and $95\%$ confidence bounds: $\mu=0\pm0.003$ (mean) and $\sigma=1\pm0.002$ (standard deviation). Note that since we normalised the curves to $\mu$ and $\sigma$, an exact fit would have been $\mu=0$ and $\sigma=1$ here. In contrast, the normalised PDFs of the $\delta v_{\perp}$ fluctuations in the ``$1/f$'' range also shown in Figure \ref{Fig.2} are clearly not Gaussian. Here they are fitted with three different heavy-tailed distributions: gamma \citep{wadsworth_97} defined by
\begin{equation}
f(x|a,b)=\frac{1}{b^a\gamma(a)}x^{a-1}e^{-x/b}
\end{equation}
with fitting parameters from maximum likelihood estimates $a=3.083\pm0.008$ and $b=0.580\pm0.002$ where the errors are from $95\%$ confidence bounds; the generalised extreme value PDF \citep{wadsworth_97} defined by
\begin{equation} f(x|k,\mu,\sigma)=\frac{1}{\sigma}exp\left(-\left(1+k\frac{(x-\mu)}{\sigma}\right)^{\frac{1}{k}}\right)\left(1+k\frac{(x-\mu)}{\sigma}\right)^{-1-\frac{1}{k}}\label{eqn7}
\end{equation}
with fitting parameters $k=0.027\pm0.001$ (shape), $\mu=0.764\pm0.001$ (location) and $\sigma=1.324\pm0.002$ (scale). The generalised extreme value (gev) distribution combines three simple extreme value distributions, types I, II and III, in a single form. The value of the shape parameter $k$ determines the type of the distribution. In the case $k\rightarrow0$, the distribution is type I, or inverse Gumbel and equation \ref{eqn7} simplifies to
\begin{equation} 
f(x|k,\mu,\sigma)=\frac{1}{\sigma}exp\left( -exp\left( -\frac{(x-\mu)}{\sigma}\right) -\frac{(x-\mu)}{\sigma}\right) \label{eqn8}
\end{equation}
This distribution corresponds to a maximum extreme value distribution or the limiting distribution of samples obtained be repeatedly selecting the maximum from an ensemble of events, which in turn, have a distribution with finite variance, e.g. Gaussian or exponential \citep{sornette_04}. Types II ($k>0$) and III ($k<0$) of the generalised extreme value distribution correspond respectively to Fr\'echet and inverse Weibull distributions. Finally a lognormal distribution \citep{wadsworth_97} defined by 
\begin{equation}
f(x|\mu,\sigma)=\frac{1}{x\sigma\sqrt{2\pi}}e^{\frac{-(lnx-\mu)^2}{2\sigma^2}}
\end{equation}
is fitted with parameters $\mu=0.410\pm0.001$ and $\sigma=0.625\pm0.001$. It can be seen from Figure \ref{Fig.4} that either the gamma distribution or the inverse Gumbel give good fits to the PDF of the $\delta v_{\perp}$ fluctuations. Physically, the gamma distribution is related to the PDF of waiting times of events generated by a Poisson process \citep{wadsworth_97}, as noted for the analysis of plasma turbulence by \citet{graves_02}.\\
Turning to the magnetic field, Figure \ref{Fig.4} shows that, unlike $\delta v_{\parallel}$, for $\delta b_{\parallel}$ there is a strong departure from the Gaussian distribution in the tails of the PDF, which are closer to stretched exponential, reminiscent of turbulence. The PDFs of the $\delta b_{\perp}$ fluctuations in the ``$1/f$'' range are fitted with the same three heavy-tailed distributions as $\delta v_{\perp}$: gamma with fitting parameters from maximum likelihood estimates $a=3.047\pm0.008$ and $b=0.585\pm0.002$ where the errors are from $95\%$ confidence bounds; generalised extreme value with fitting parameters $k=0.017\pm0.002$, $\mu=0.770\pm0.001$ and $\sigma=1.325\pm0.002$; and lognormal with fitting parameters $\mu=0.406\pm0.001$ and $\sigma=0.633\pm0.001$. There is little difference between the raw and collapsed PDFs, as $\delta v=b_{\parallel,\perp}$ is closer to ``$1/f$'' scaling. 
\begin{figure}[H]
\figurenum{5}
\epsscale{0.4}
\plotone{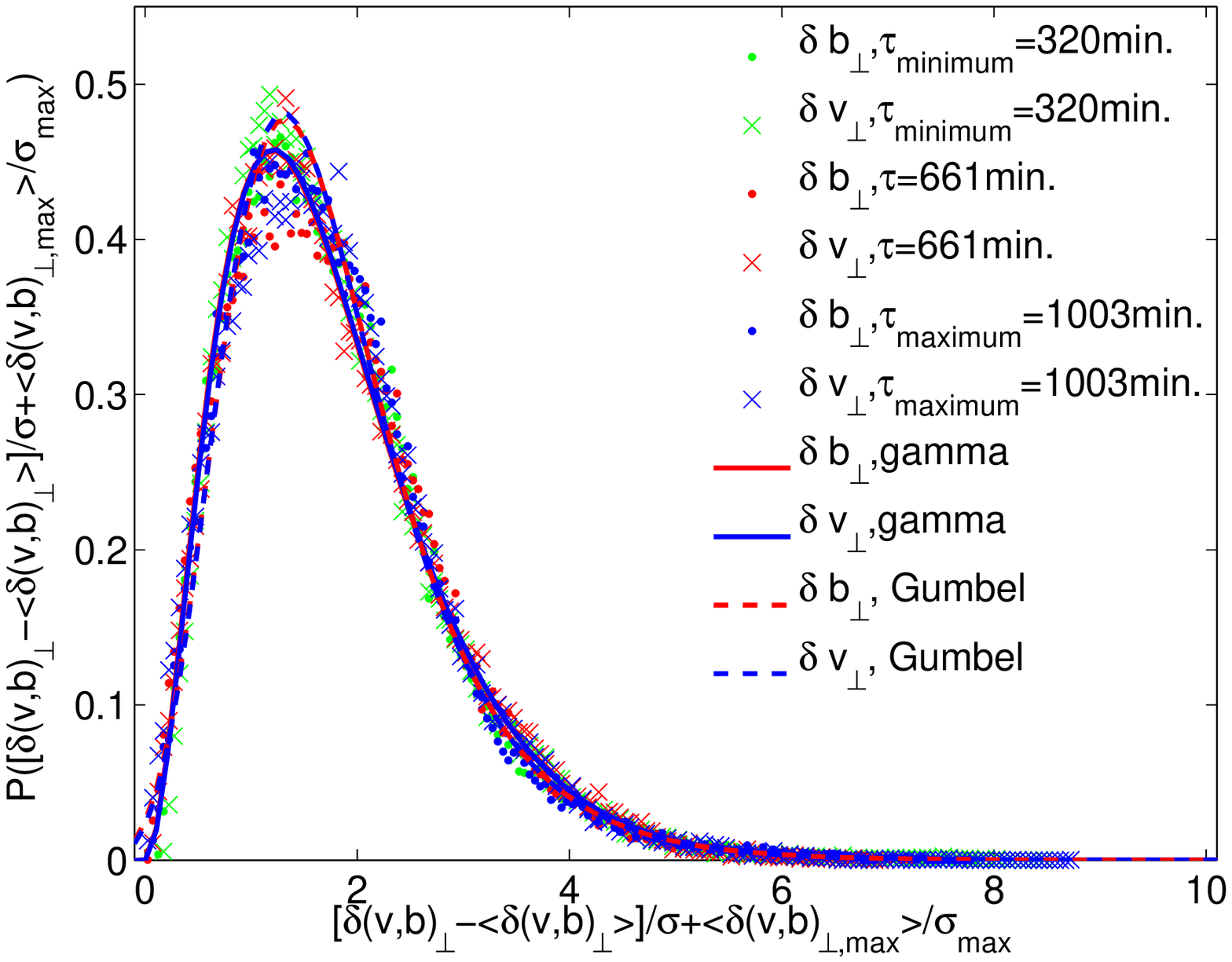}
\plotone{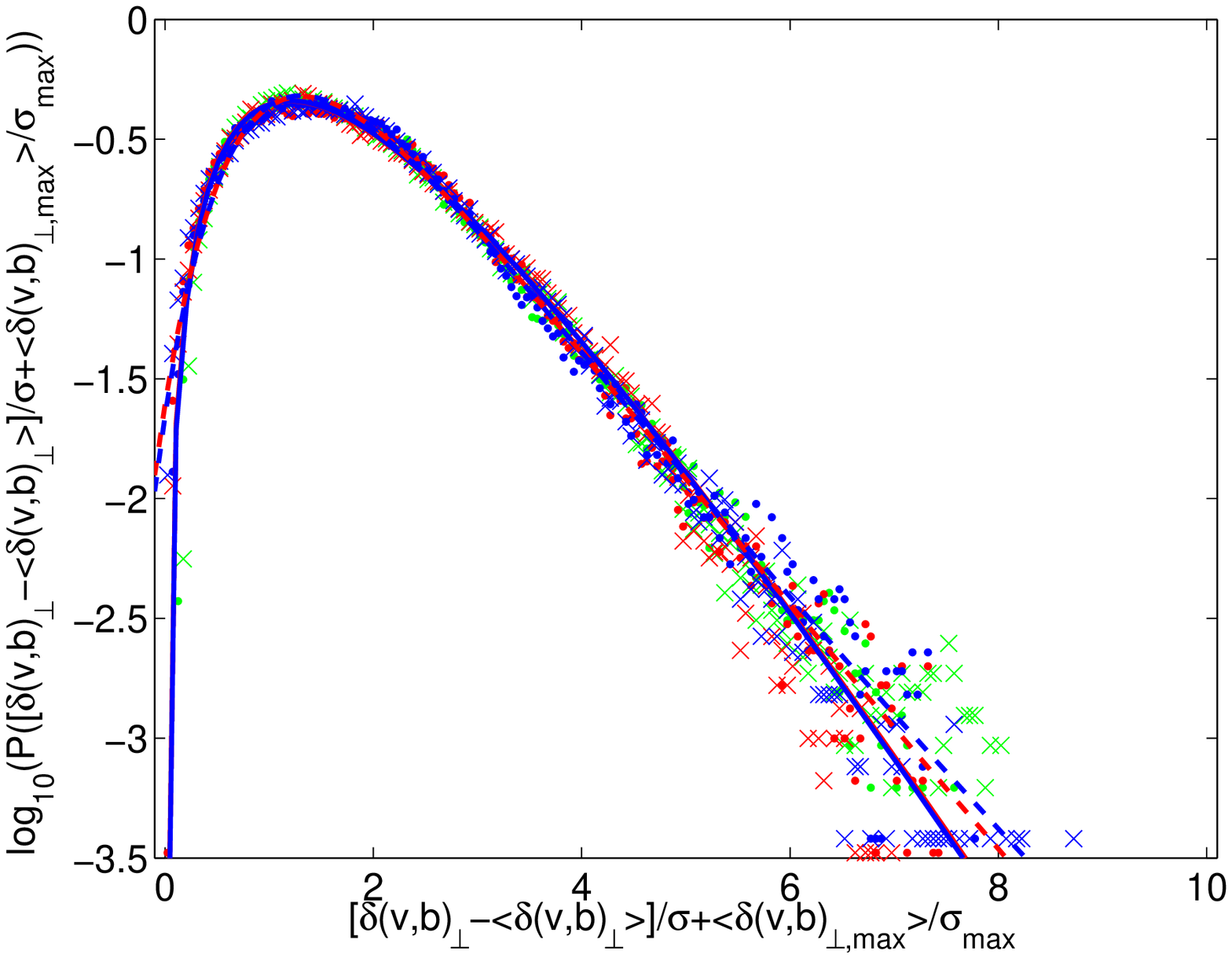}
\caption{Perpendicular magnetic field fluctuations $\delta b_{\perp}$ (``$\cdot$'') and $\delta v_{\perp}$ (``$\times$'') in the fast solar wind at solar minimum for the ``$1/f$'' range. The gamma and Gumbel distributions are used to fit the normalised curves for $\delta b_{\perp}$ (red) and $\delta v_{\perp}$ (blue). Only three representative values of $\tau$ are shown for each component, whereas the fitted curves are computed using all the $\tau$ intervals. The left panel shows the PDFs on linear axis, whereas the right panel shows the normalised curves on a semilog $y$ axis.}
\label{Fig.5}
\end{figure}
As we have seen, $\delta b_{\perp}$ and $\delta v_{\perp}$ appear to be strongly similar in their statistics and Figure \ref{Fig.5} overlays the normalised PDFs for $\delta b_{\perp}$ and $\delta v_{\perp}$ in the fast quiet solar wind. We see that they are almost identical. A possible interpretation is that both sets of fluctuations have the same physical process at their origin.\\For completeness, we also examine the ion density fluctuations $\delta\rho$ in the fast quiet solar wind. From \citet{matthaeus_07}, one might expect these to show similar scaling behaviour to the $\delta b_{\parallel,\perp}$ fluctuations, however in Figure \ref{Fig.6} we see that this is not the case. The density PDFs have very sharp peaks with extended tails and are asymmetric. The rescaling collapse works well at the centre of the PDFs, but not towards the tails.
\begin{figure}[H]
\figurenum{6}
\epsscale{0.4}
\plotone{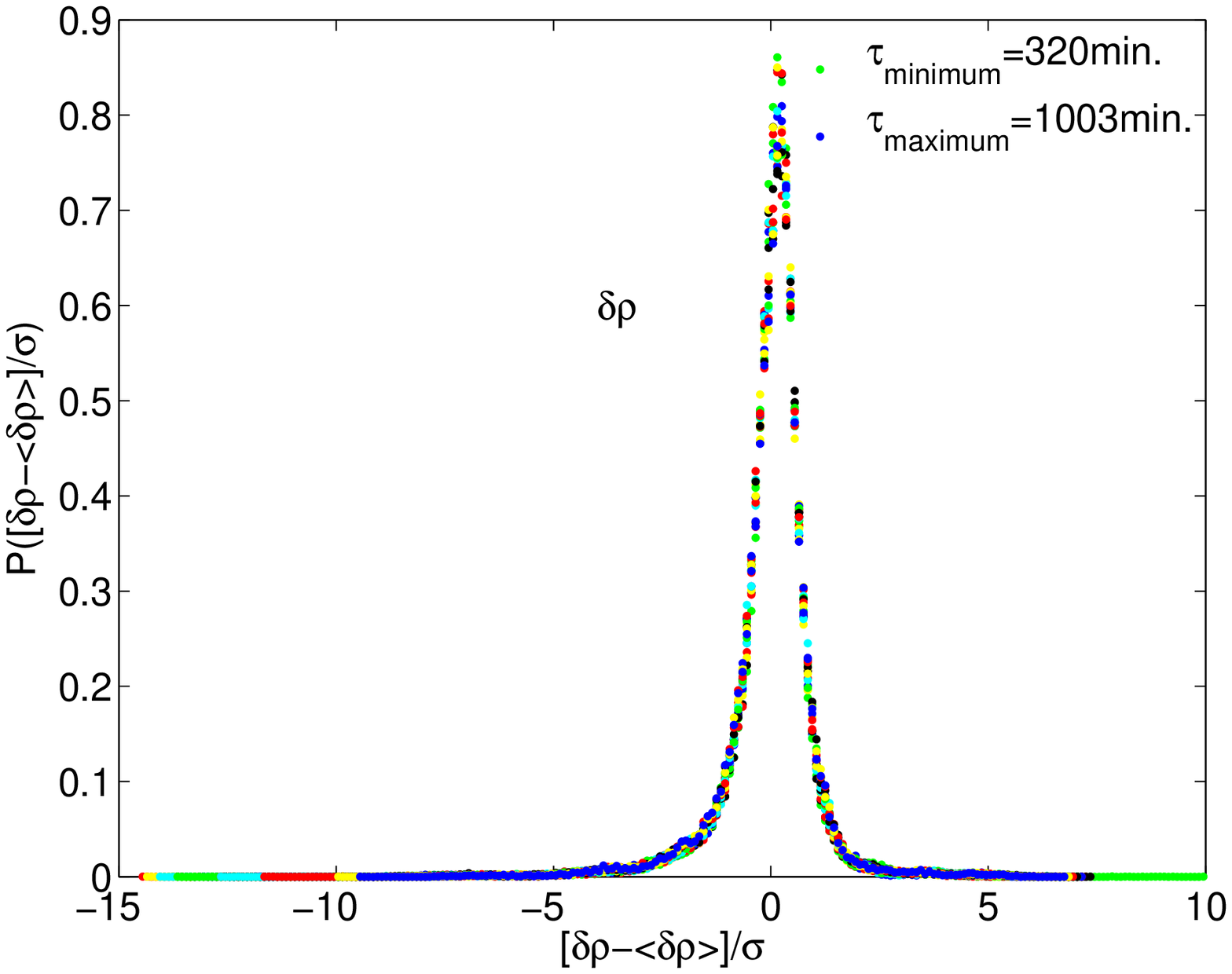}
\plotone{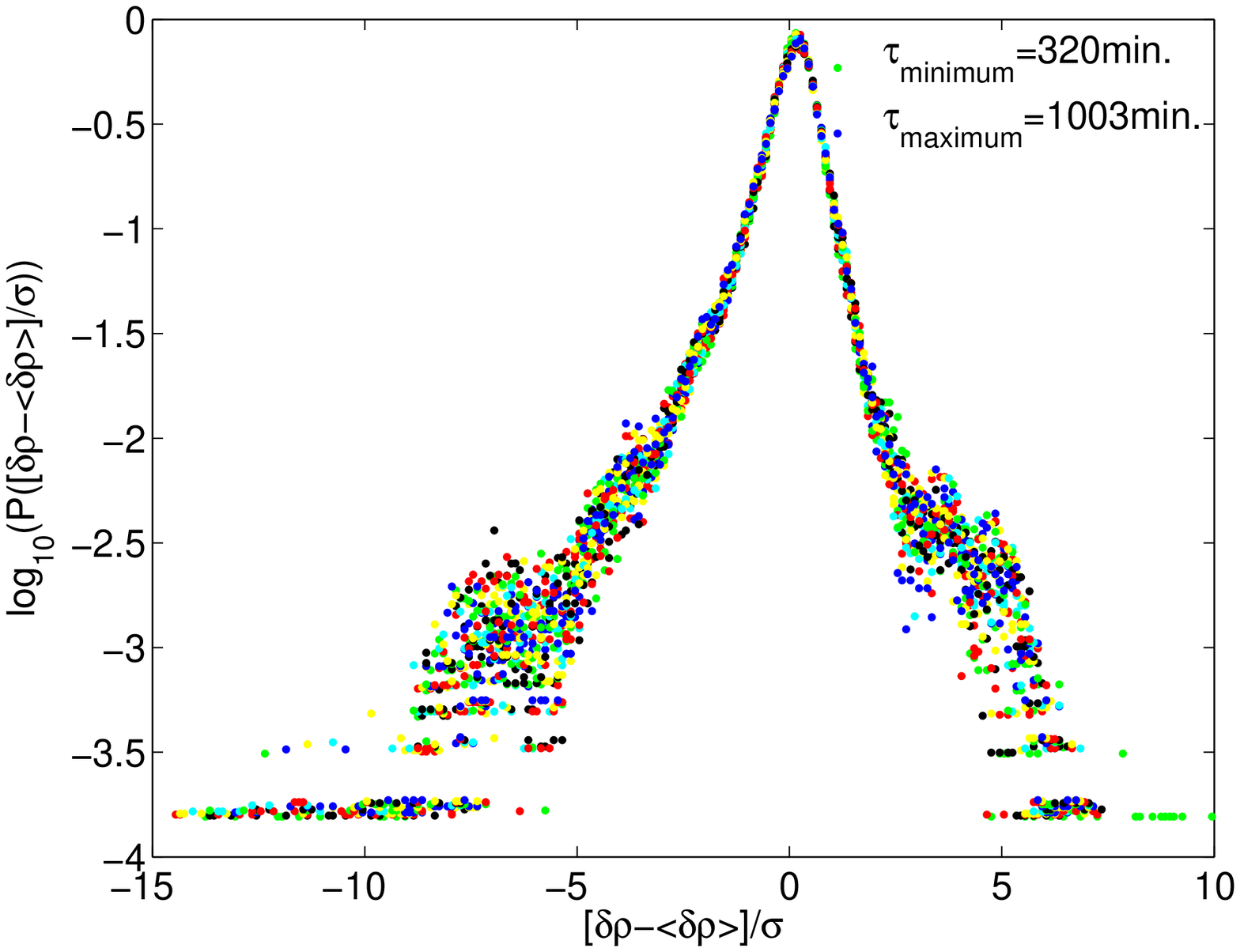}
\caption{Ion density fluctuations $\delta \rho$ in the fast solar wind at solar minimum for the ``$1/f$'' range. The left panel shows the PDFs of raw fluctuations sampled across intervals $\tau$ between $320$ to $1003$ minutes normalised using equation \ref{eqn6}. The right panel shows the normalised curves on a semilog $y$ axis.}
\label{Fig.6}
\end{figure}
To conclude this section, let us summarise our analysis of the PDFs of fluctuations in the fast quiet solar wind. Figures \ref{Fig.2} and \ref{Fig.3} (right-hand sides) show scaling collapse for $\delta v_{\parallel,\perp}$ and $\delta b_{\parallel,\perp}$. Figure \ref{Fig.4} (left-hand side) shows that $\delta v_{\parallel}$ and $\delta b_{\parallel}$ are distinct. This is manifest in both a different functional form of the rescaled PDFs and different scaling of the moments, which we discuss next. In particular, $\delta b_{\parallel}$ is nearly symmetric and has stretched exponential tails, consistent with a multiplicative process, whereas $\delta v_{\parallel}$ is more asymmetric and is close to Gaussian. Figure \ref{Fig.5} shows that $\delta v_{\perp}$ and $\delta b_{\perp}$ have the same PDF functional form and are reasonably well fitted by the gamma and Gumbel distributions with similar fitting parameters, suggesting a common source for the fluctuations.
\section{GSF analysis for comparison of quiet fast and slow streams}
\begin{figure}[H]
\figurenum{7}
\epsscale{0.4}
\plotone{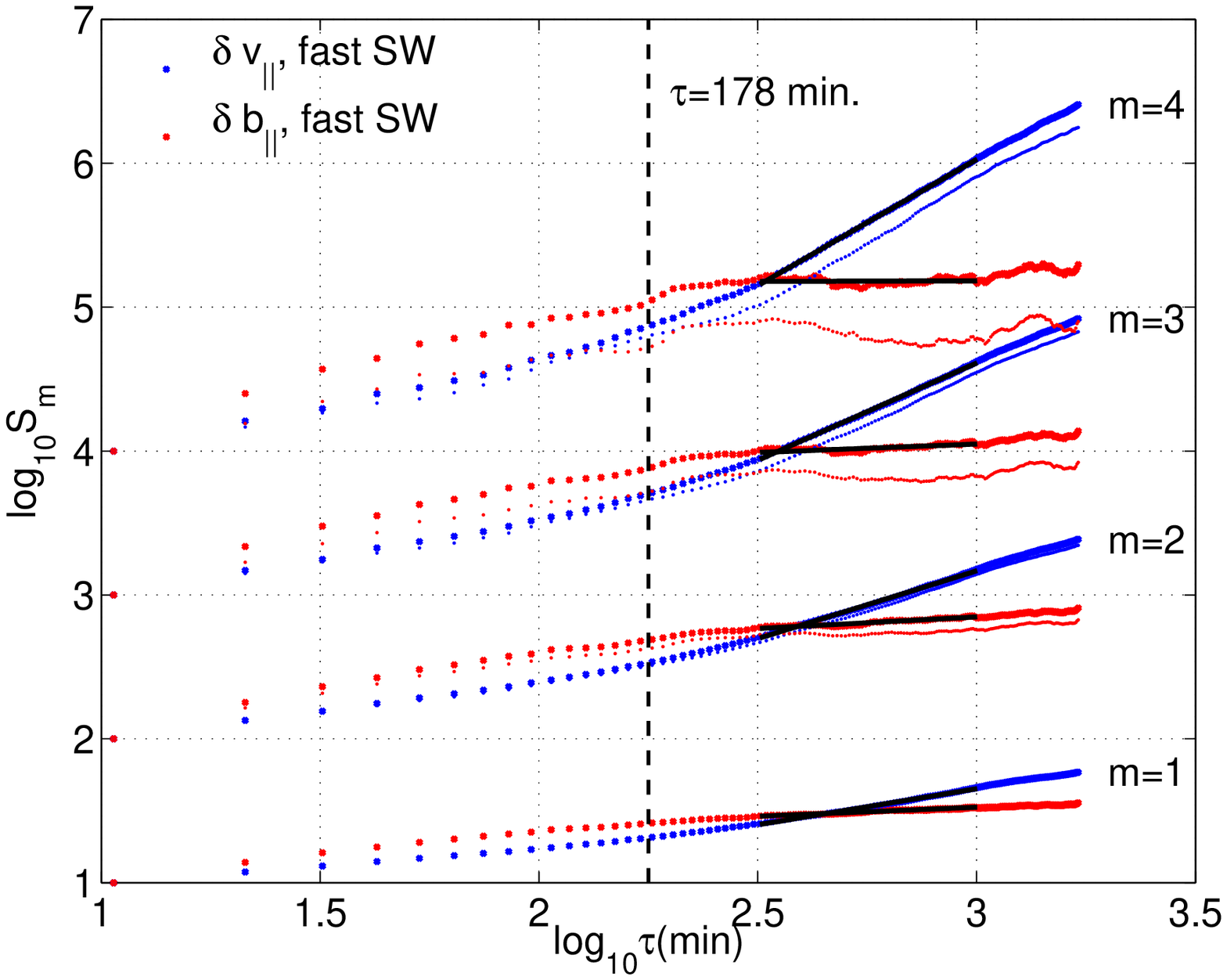}
\plotone{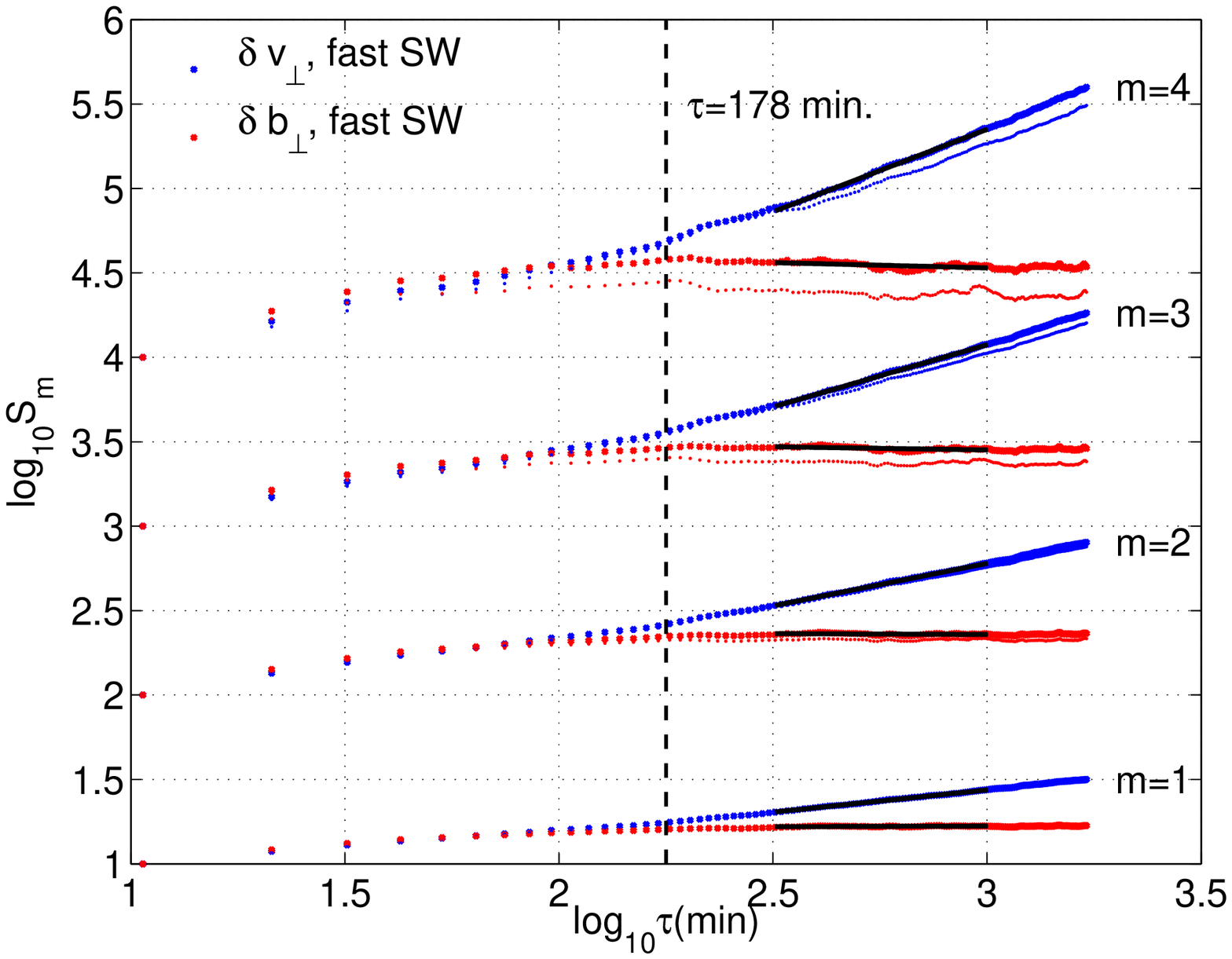}
\plotone{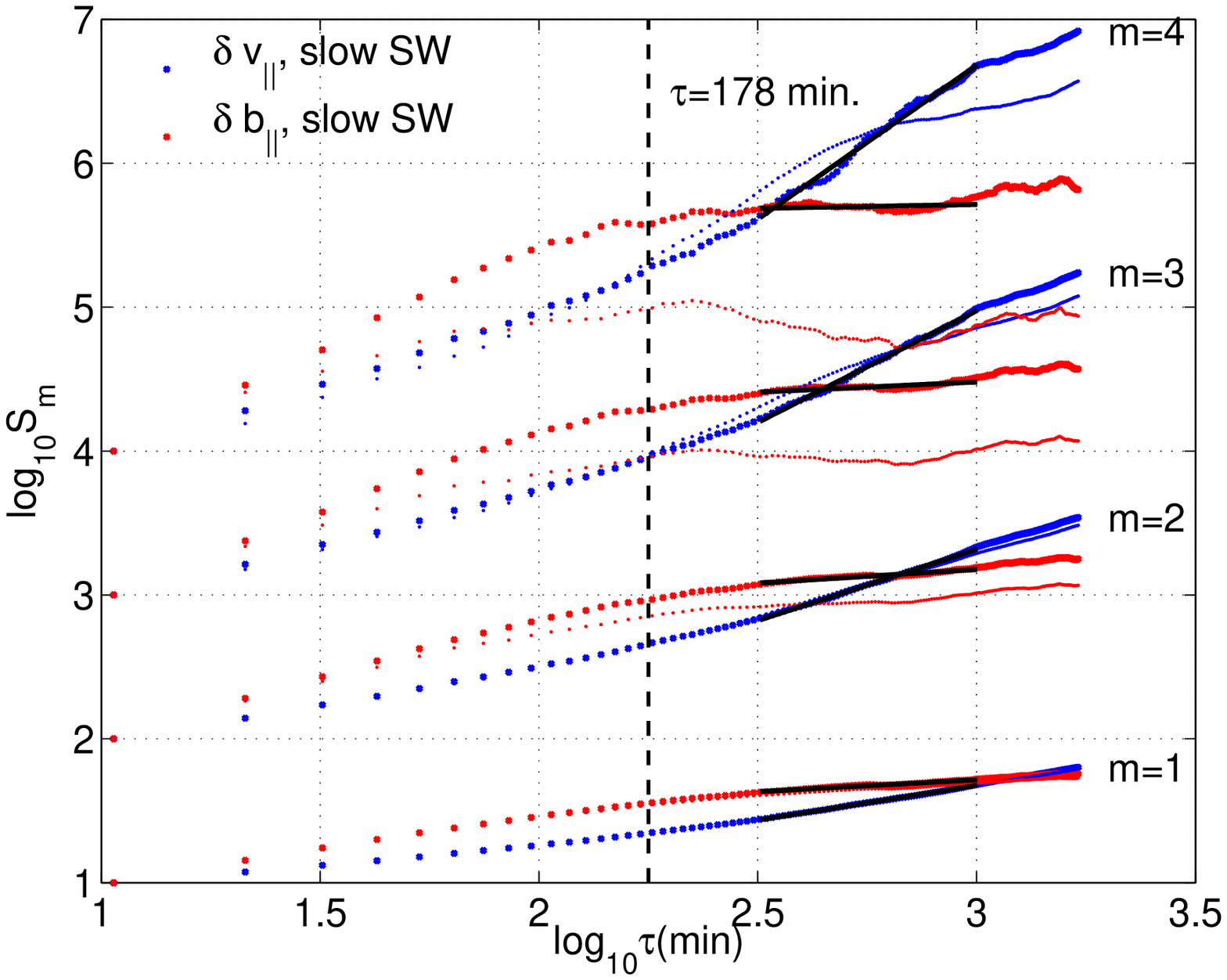}
\plotone{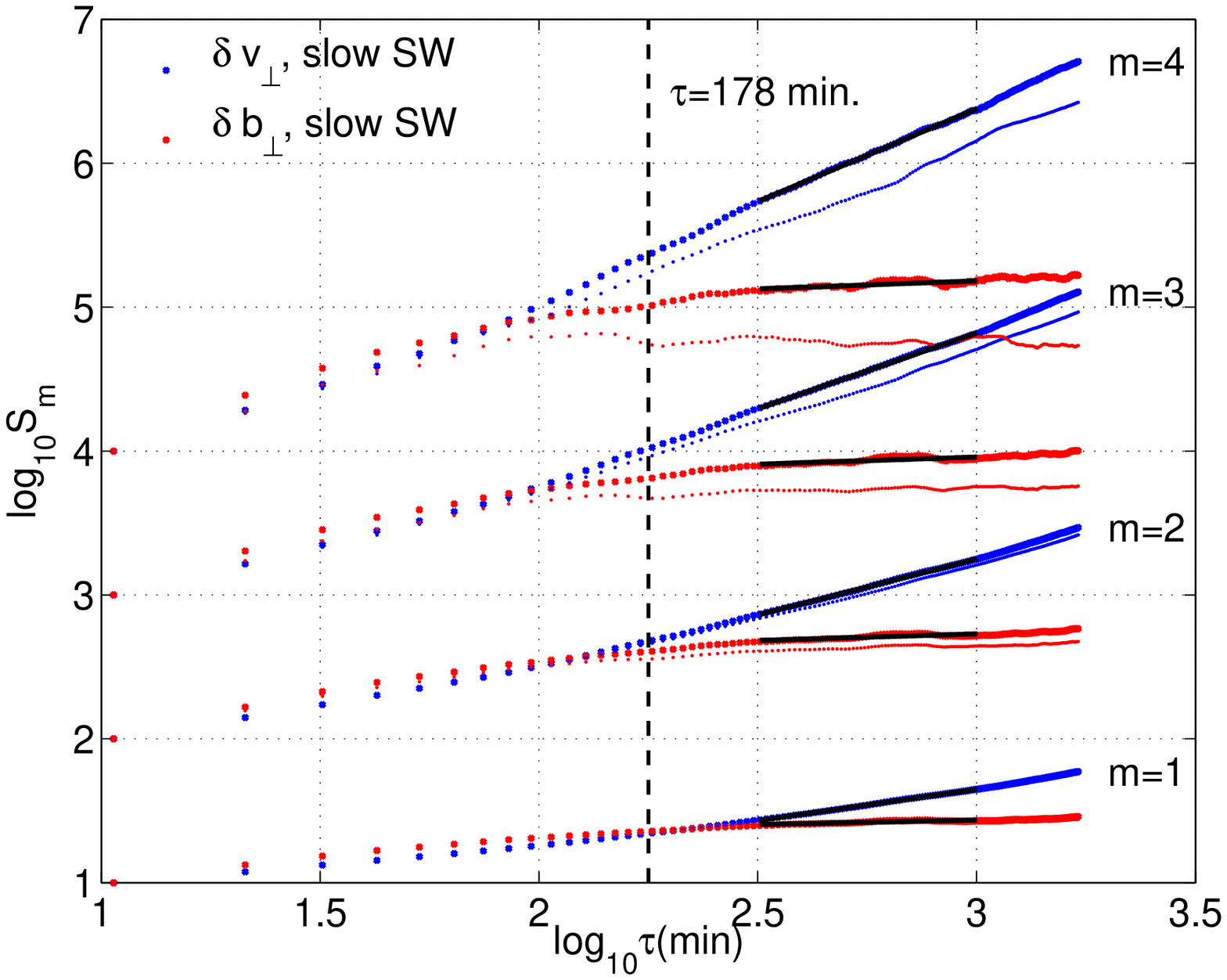}
\caption{Comparison of scaling properties of fluctuations in the fast solar wind (upper) and in the slow solar wind (lower) at solar minimum in $2007$. Generalised structure functions $S_{m}$ are plotted on log-log axes versus sampling interval $\tau$ for $\tau=10$ to $1706$ minutes and $m=1$ to $4$. Left panels show parallel components of fluctuations in velocity (blue) and magnetic field (red); right panels show corresponding perpendicular components. The raw data is shown by ``$\cdotp$'', whereas ``\mbox{\boldmath\large$\cdotp$}'' denote data which has been conditioned by $0.4\%$ (the difference in these curves quantifies finite size effects). Linear regression fits to the ``$1/f$'' range over $\tau=320$ to $1002$ minutes are shown. The transition from the IR to the ``$1/f$'' range occurs at $\sim178$ minutes and is shown by the dashed line.}
\label{Fig.7}
\end{figure}

\begin{figure}[H]
\figurenum{8}
\epsscale{0.4}
\plotone{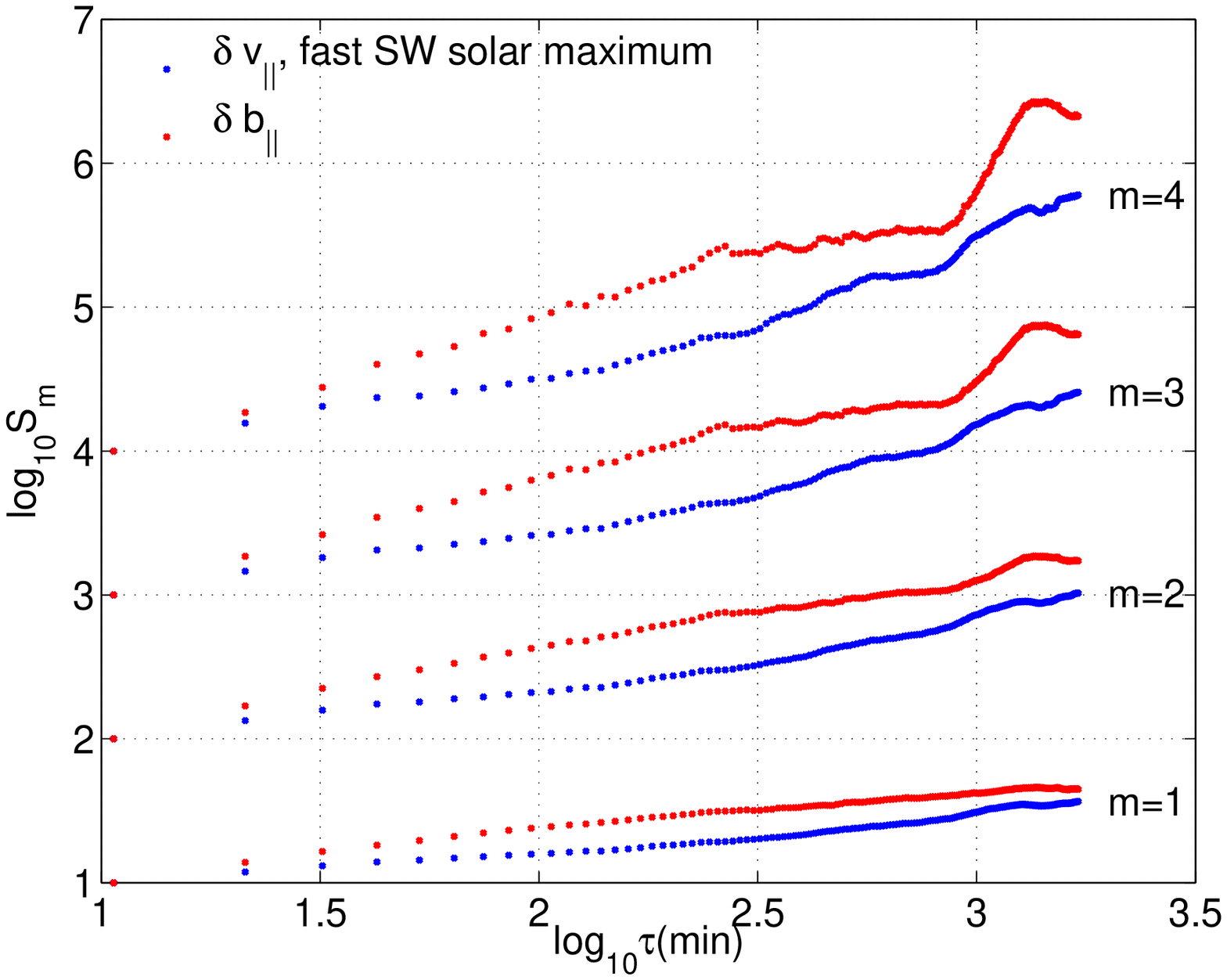}
\plotone{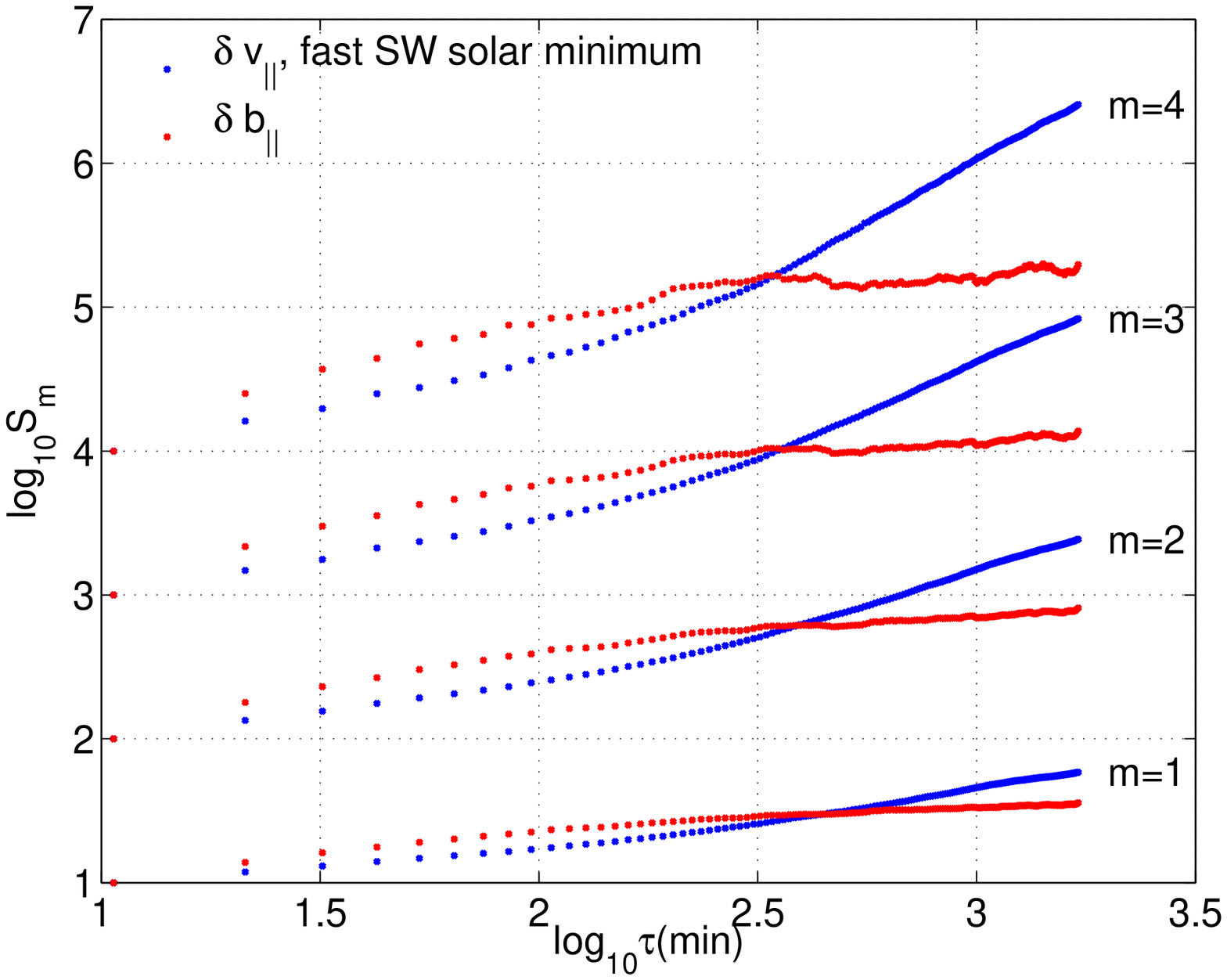}
\plotone{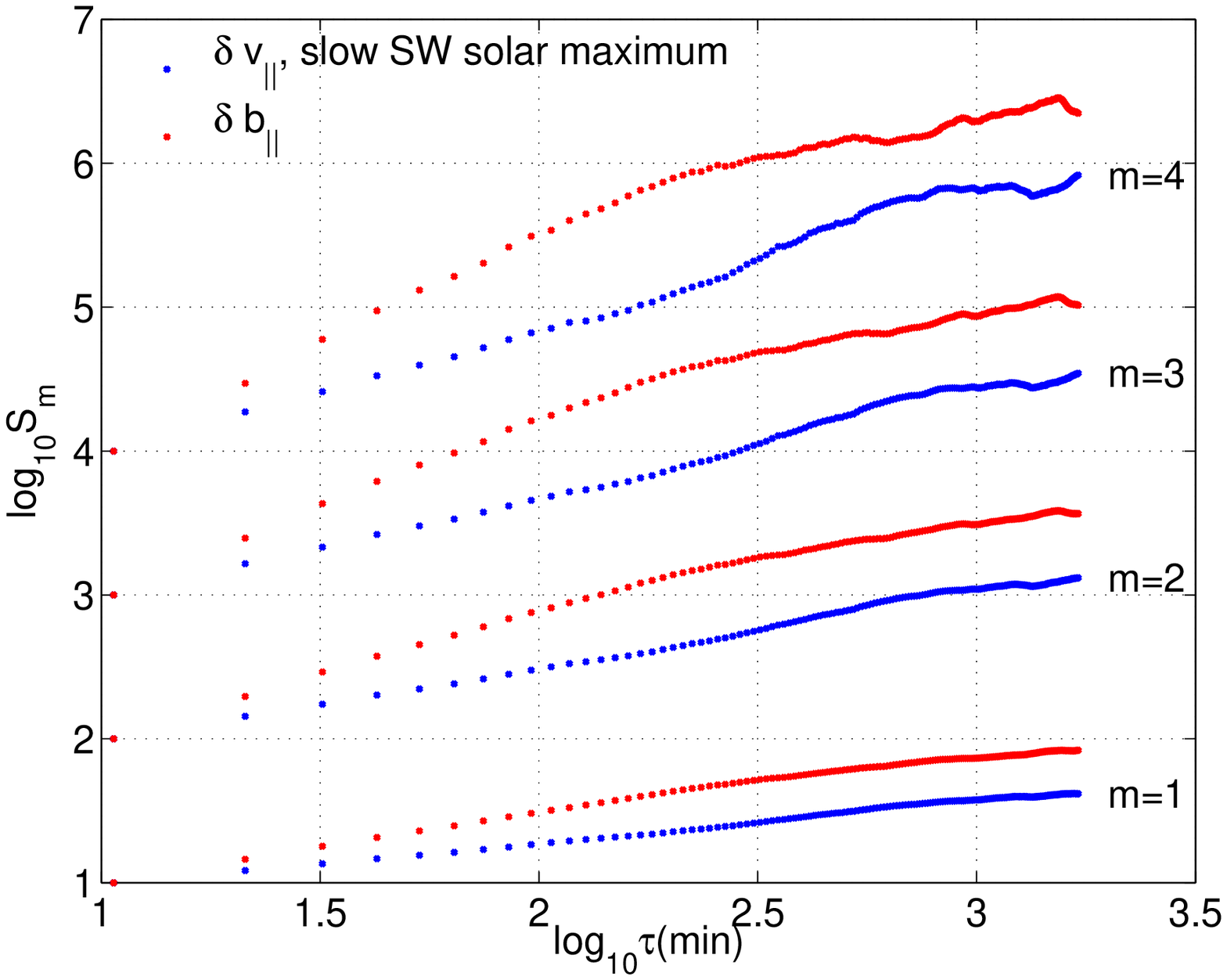}
\plotone{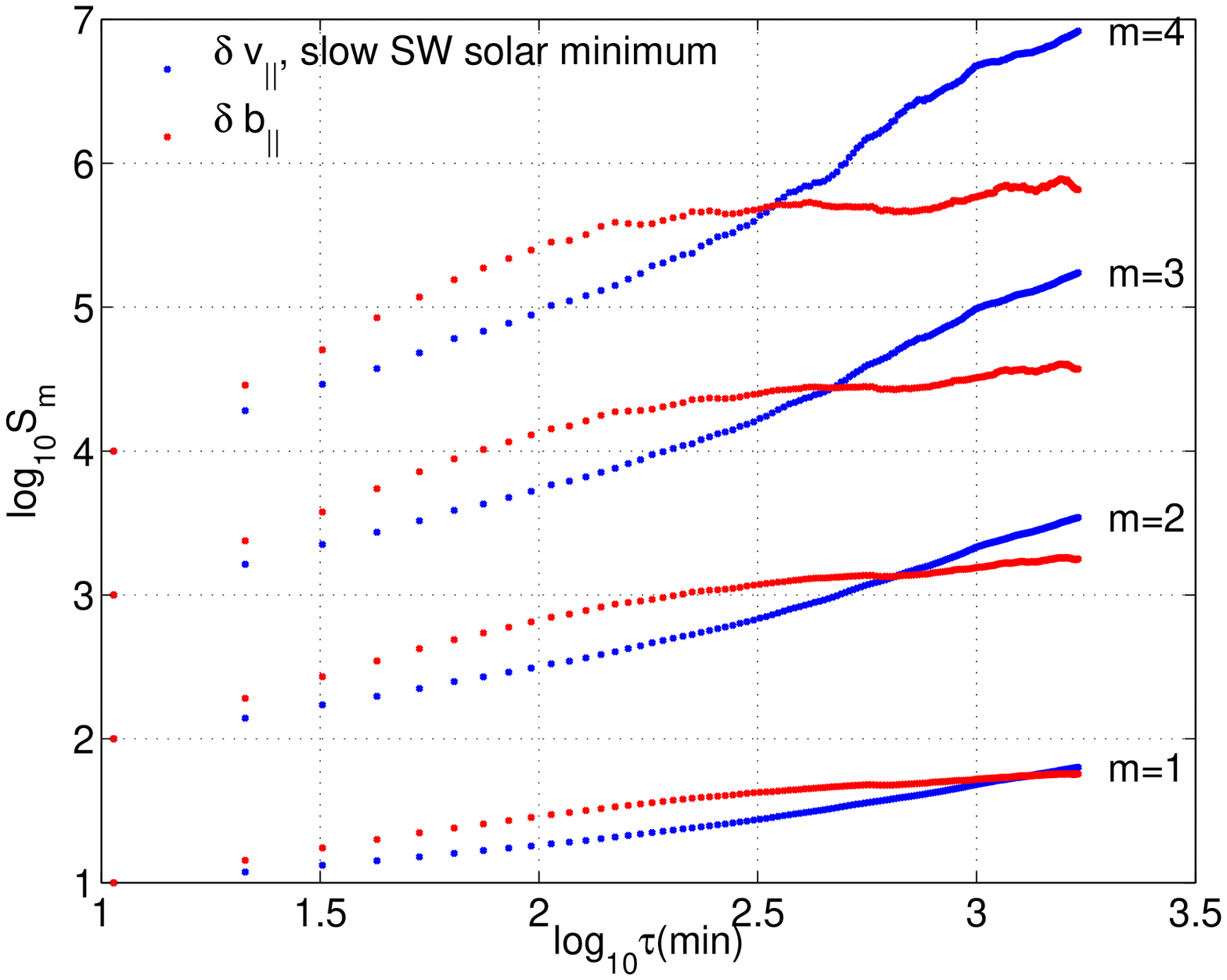}
\caption{Comparison of scaling properties of parallel fluctuations between fast (upper) and slow (lower) solar wind at solar maximum (left) in $2000$ and solar minimum (right) in $2007$. Generalised structure functions $S_{m}$ are plotted on log-log axes versus sampling interval $\tau$ for $\tau=10$ to $1706$ minutes and $m=1$ to $4$. Parallel components of fluctuations in velocity (blue) and magnetic field (red) are shown.}
\label{Fig.8}
\end{figure}

\begin{figure}[H]
\figurenum{9}
\epsscale{0.4}
\plotone{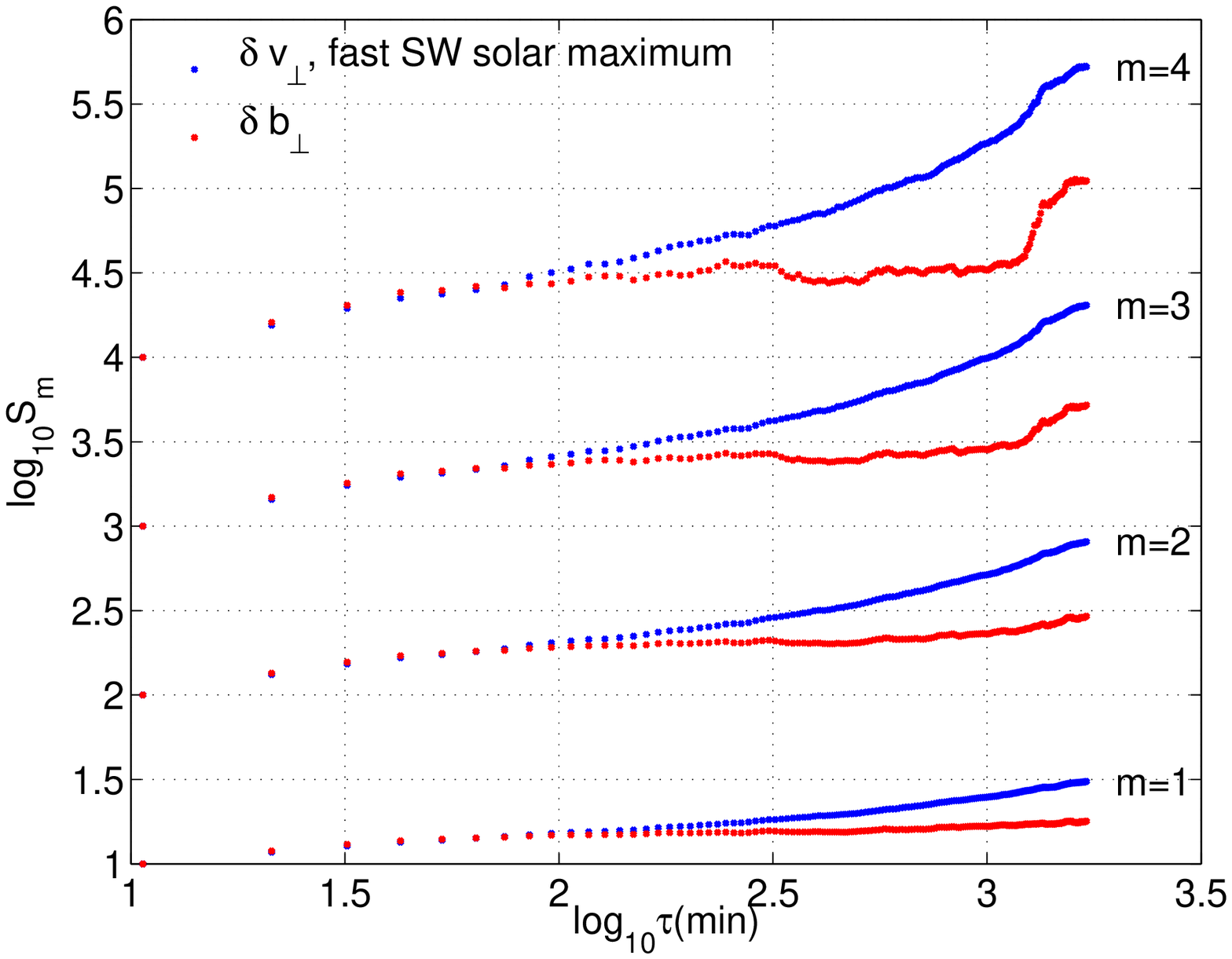}
\plotone{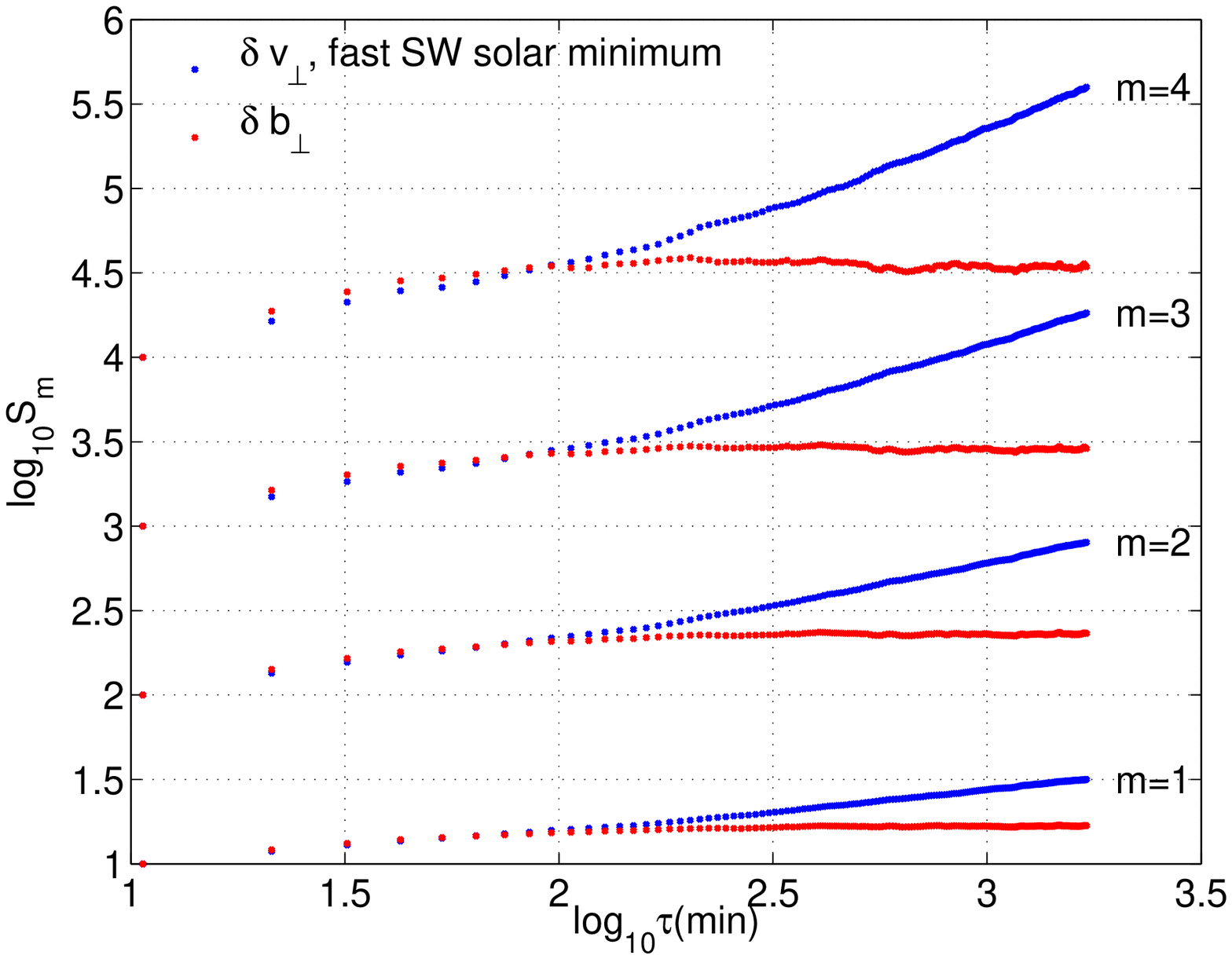}
\plotone{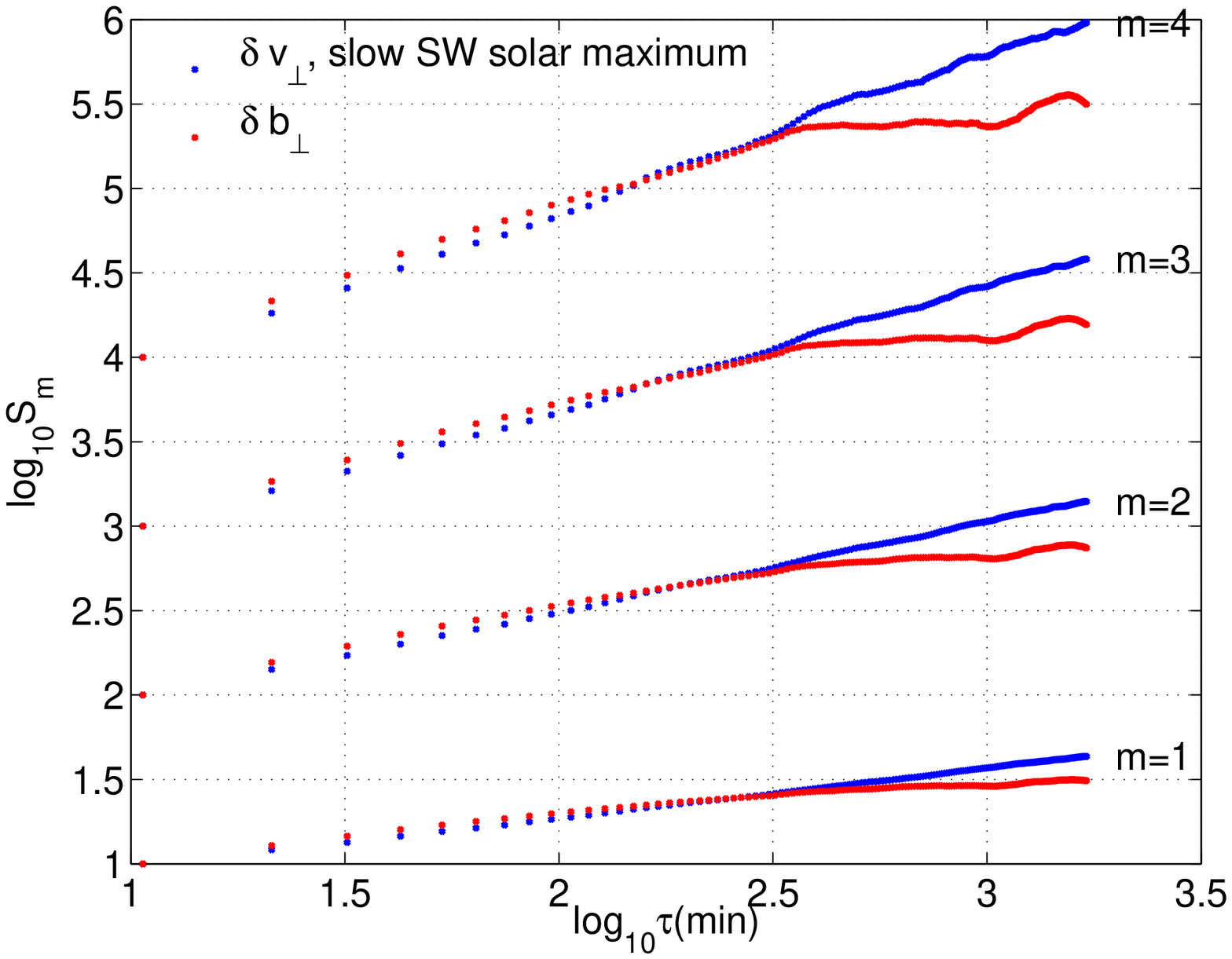}
\plotone{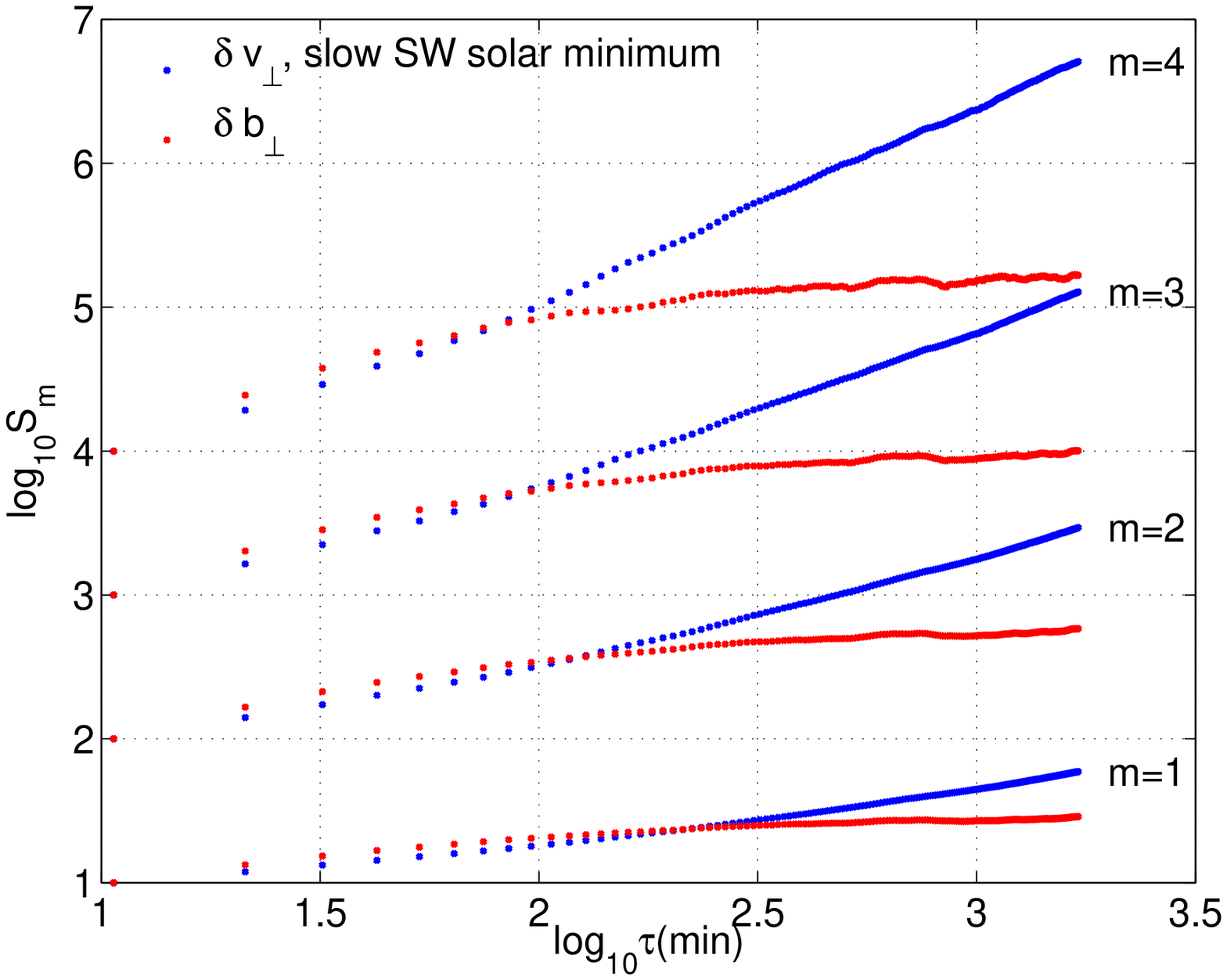}
\caption{Comparison of scaling properties of perpendicular fluctuations between fast (upper) and slow (lower) solar wind at solar maximum (left) in $2000$ and solar minimum (right) in $2007$. Generalised structure functions $S_{m}$ are plotted on log-log axes versus sampling interval $\tau$ for $\tau=10$ to $1706$ minutes and $m=1$ to $4$. Perpendicular components of fluctuations in velocity (blue) and magnetic field (red) are shown.}
\label{Fig.9}
\end{figure}

Scaling can be quantified by computing the generalised structure functions (GSFs) of the fluctuations, $\langle\mid\delta y_{\tau}\mid^{m}\rangle$, where $\langle\cdots\rangle$ denotes ensemble averaging, $m$ is the order of the moment and $\delta y_{\tau}=y(t+\tau)-y(t)$ is the fluctuation in a signal $y(t)$ over a time $\tau$. Assuming weak stationarity and a degree of self-similarity, GSFs can be related to the scale $\tau$ of the fluctuation by a scaling exponent $\zeta(m)$, when
\begin{equation}
S_{m}=\langle\mid\delta y_{\tau}\mid^{m}\rangle\sim \tau^{\zeta(m)}\label{eqn4}
\end{equation}
whereas the PSD measures $\zeta(2)$ only \citep[e.g.][]{marsch_97,horbury_97}. We anticipate scaling for the datasets considered here, given the indication of a ``$1/f$'' range in the PSDs in Figure \ref{Fig.1}, however power spectra alone cannot distinguish between fractal and multifractal behaviour \citep{chapman_05}. From equation (\ref{eqn4}), the scaling exponents $\zeta(m)$ are given quantitatively by the slopes of the GSFs. Generally for perfectly self-affine processes, $\zeta(m)$ can be described by a linear equation $\zeta(m)=Hm$, where $H$ is the Hurst exponent. Each successful computation of a GSF at increasingly high order yields additional information about the nature of the PDF of fluctuations. For practical applications of the GSF analysis to a broad range of datasets, see for example: MHD turbulence simulations -  \citet{merrifield_05,merrifield_06,merrifield_07}; solar wind turbulence - \citet{horbury_97,hnat_05b,chapman_07,nicol_08}; geomagnetic indices \citet{hnat_03b}; laboratory plasma turbulence - \citet{budaev_06,dewhurst_08,hnat_08} and references therein.\\
We now apply these methods to the observations. Figure \ref{Fig.7} shows the GSFs up to $m=4$ for $\delta v_{\parallel}$, $\delta v_{\perp}$, $\delta b_{\parallel}$ and $\delta b_{\perp}$ for fast and slow solar wind at solar minimum. The series is differenced over $\tau=n\times640$ s for $n=1$ to $160$, that is for a range covering ten to $1706$ minutes ($\sim28$ hours). The finite length of the datasets considered means that the statistics calculated for any given single ensemble can in principle be affected by the presence of large outliers, which are insufficiently numerous to be fully sampled. We check that this does not bias our results via the method of \citet{kiyani_07}, which by subtracting outliers verifies whether calculated exponents are robust against statistical fluctuations in the outliers. The raw and $0.4\%$ conditioned GSFs are shown for comparison in Figure \ref{Fig.7}. For the low-order moments that we consider here, we see that the difference is small, so that the finite length of our datasets does not significantly affect our conclusions. The raw data is used for the plots of the probability densities of the fluctuations in section $3$.
Figure \ref{Fig.7} is consistent with the results shown previously, namely that $\mbox{\boldmath$v$}$ and $\mbox{\boldmath$B$}$ fluctuations exhibit very different behaviour in the ``$1/f$'' range, which corresponds to large $\tau$ intervals. A simple self-affine noise process with PSD$\sim1/f^{\alpha}$, $\alpha\sim1$ would on such a plot have $\zeta(2)\rightarrow0$ since $\alpha=1+\zeta(2)$. If the process is fractal then $\zeta(m)=\alpha m\rightarrow0$ for all $m$. Thus we see that at $\tau>178$ min., the GSFs for $\delta b_{\parallel,\perp}$ ``flatten'' in the $\sim1/f$ range, consistent with previously reported results based on the PSD \citep{matthaeus_86,matthaeus_07}. The $\delta v_{\parallel,\perp}$ GSFs on the contrary steepen at $\tau>178$ minutes, showing a scaling process and exponents distinct from those of $\delta b_{\parallel,\perp}$. These are closer to a value of $\zeta(2)\sim1$, which, again for a simple noise process, is consistent with PSD$\sim1/f^{2}$. This is what we have seen in the PDF curve renormalization of the previous section: the $\delta b_{\parallel,\perp}$ raw PDFs were close to the renormalised PDFs, since the normalization is with respect to the first two moments $S_{1}$ and $S_{2}$, which for $\delta b_{\parallel,\perp}$ vary weakly as a function of scale $\tau$. Equation \ref{eqn4} tells one that the scaling behaviour of the process is contained in the $\zeta(m)$ exponents, given by the slopes of the GSFs. We obtain these values by linear fits to the log-log GSF plots.\\
Whilst these results confirm the ``$1/f$'' scaling of fluctuations in $\mbox{\boldmath$B$}$ on long timescales, reported previously by for example \citet{matthaeus_86}, they also highlight the distinct scaling of $\mbox{\boldmath$v$}$, which we will investigate next. These GSF plots of fluctuations oriented with respect to the background field also clearly show the crossover between the IR and ``$1/f$'' for fast and slow solar wind. The ``$1/f$'' range is much shorter in the slow streams, consistent with previous observations \citep[e.g.][]{bruno_05,horbury_05}. The minimum value of $\tau$ that we will use for the following analysis can be seen to be greater than the breakpoint $\tau$ for both velocity and magnetic field fluctuations. For $\delta v_{\parallel}$ and $\delta b_{\parallel}$ in both fast and slow wind, the timescale $\tau$ at which the GSFs diverge is $\gtrsim\tau=128$ minutes, the spectrally inferred breakpoint between IR and ``$1/f$''. In contrast, the divergence between the GSFs of $\delta v_{\perp}$ and $\delta b_{\perp}$ begins at a significantly shorter timescale $\tau\sim100$ minutes. This is particularly apparent when one considers the higher order moments such as $m=3,\;4$ in Figure \ref{Fig.7}. It is also interesting to note that although the PDFs of $\delta v_{\perp}$ and $\delta b_{\perp}$ in the fast quiet solar wind show the same functional form (Figure \ref{Fig.5}), their GSF scalings are very different. This   may suggest that the fluctuations $\delta v_{\perp}$ and $\delta b_{\perp}$ originate in a common coronal source, but their subsequent development differs in the evolving and expanding solar wind.\\
Figures \ref{Fig.8} and \ref{Fig.9} compare the GSFs for fast and slow solar wind streams at solar maximum ($2000$) and minimum ($2007$); for clarity only the $0.4\%$ conditioned results are shown.
Figures \ref{Fig.8} and \ref{Fig.9} suggest that the scaling properties of $\delta v_{\perp}$ and $\delta b_{\perp}$ do not change with solar cycle in fast solar wind. However $\delta v_{\parallel}$ does, while the solar cycle dependence of $\delta b_{\parallel}$ is indeterminate. The divergences at $\tau\gtrsim10^3$ in the fast solar wind at solar maximum may be due to finite size effects: the dataset for solar maximum is shorter than for solar minimum, because there are fewer long continuous time periods of fast solar wind. Figures \ref{Fig.8} and \ref{Fig.9} also show that the scaling properties of all four fluctuating quantities in the slow solar wind differ between solar maximum and minimum, due to different scaling exponents or a different breakpoint location.\\
Let us summarise our conclusions from the GSF analysis. First, the breakpoints between the scaling properties measured by GSF analysis are different between fast and slow solar wind streams, and between periods of maximum and minimum solar activity. These breakpoints do not necessarily coincide with the breakpoint between IR and ``$1/f$'' ranges inferred from spectral analysis, however as mentioned earlier, it is difficult to extract precise quantitative information from the power spectra plots. The IR extends to longer timescales in slow solar wind streams and at periods of maximum solar activity \citep[e.g.][]{horbury_05}. These trends are particularly clear in the GSFs of the perpendicular components in Figure \ref{Fig.9}. The inertial range remains relatively robust for both slow and fast solar wind streams and is independent of solar cycle. This is to be expected if the IR is established by a turbulent cascade within the evolving expanding solar wind, rather than by initial conditions in the corona. Intriguingly, $\delta v_{\perp}$ and $\delta b_{\perp}$ have the same behaviour in the ``$1/f$'' range for fast solar wind at both solar maximum and minimum. Their scaling looks similar for the slow solar wind, but the breakpoint moves to longer timescales at solar maximum. All four quantities vary between fast and slow solar wind and solar minimum and maximum.
\section{Quantifying the scaling exponents}
Let us now quantify the observed scaling by measuring the slopes of the GSFs to obtain estimates of the values of the scaling exponents, $\zeta(m)$; the robustness of the scaling will also be tested. In principle, values for $\zeta(m)$ are obtained from the gradients of the log-log plots of $S_{m}$ versus $\tau$. In practice, these are affected by the fact that both the length of the dataset, and the range of $\tau$ over which we see scaling, are finite. As a preliminary, therefore, we outline a method to optimise this process to obtain a good estimate of the exponents.
\subsection{p-model and Brownian walk test timeseries}
\begin{figure}[H]
\figurenum{10}
\epsscale{0.4}
\plotone{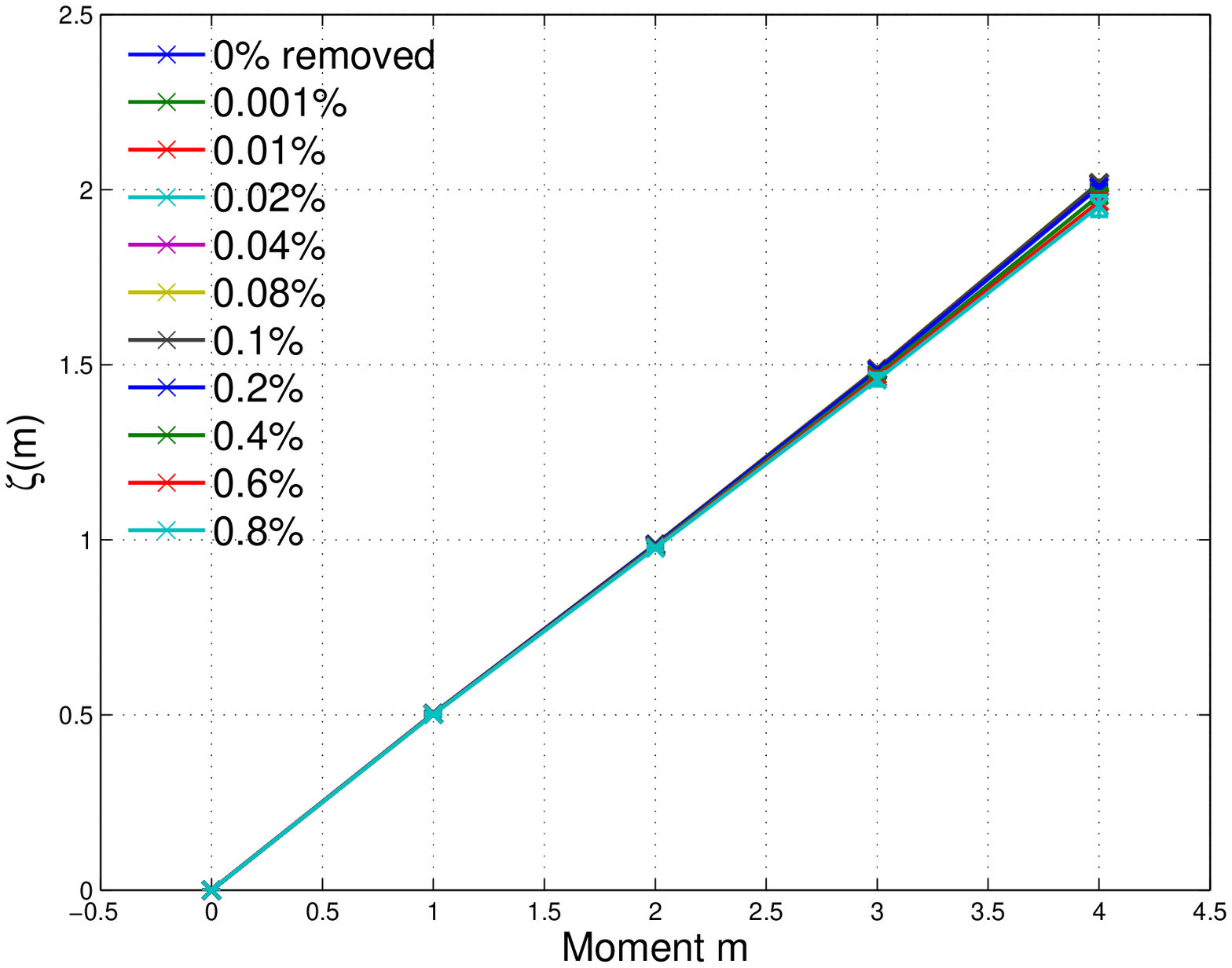}
\plotone{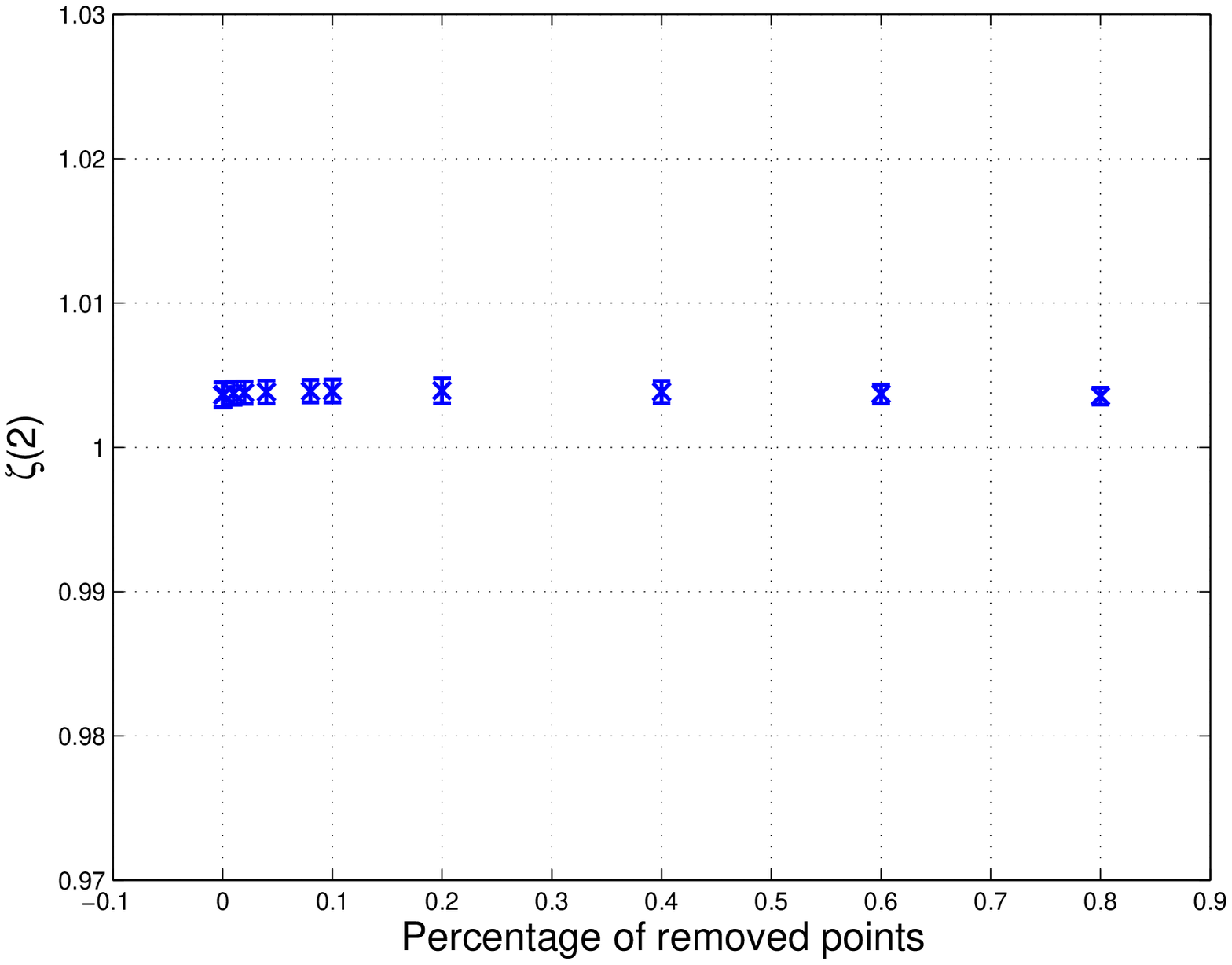}
\plotone{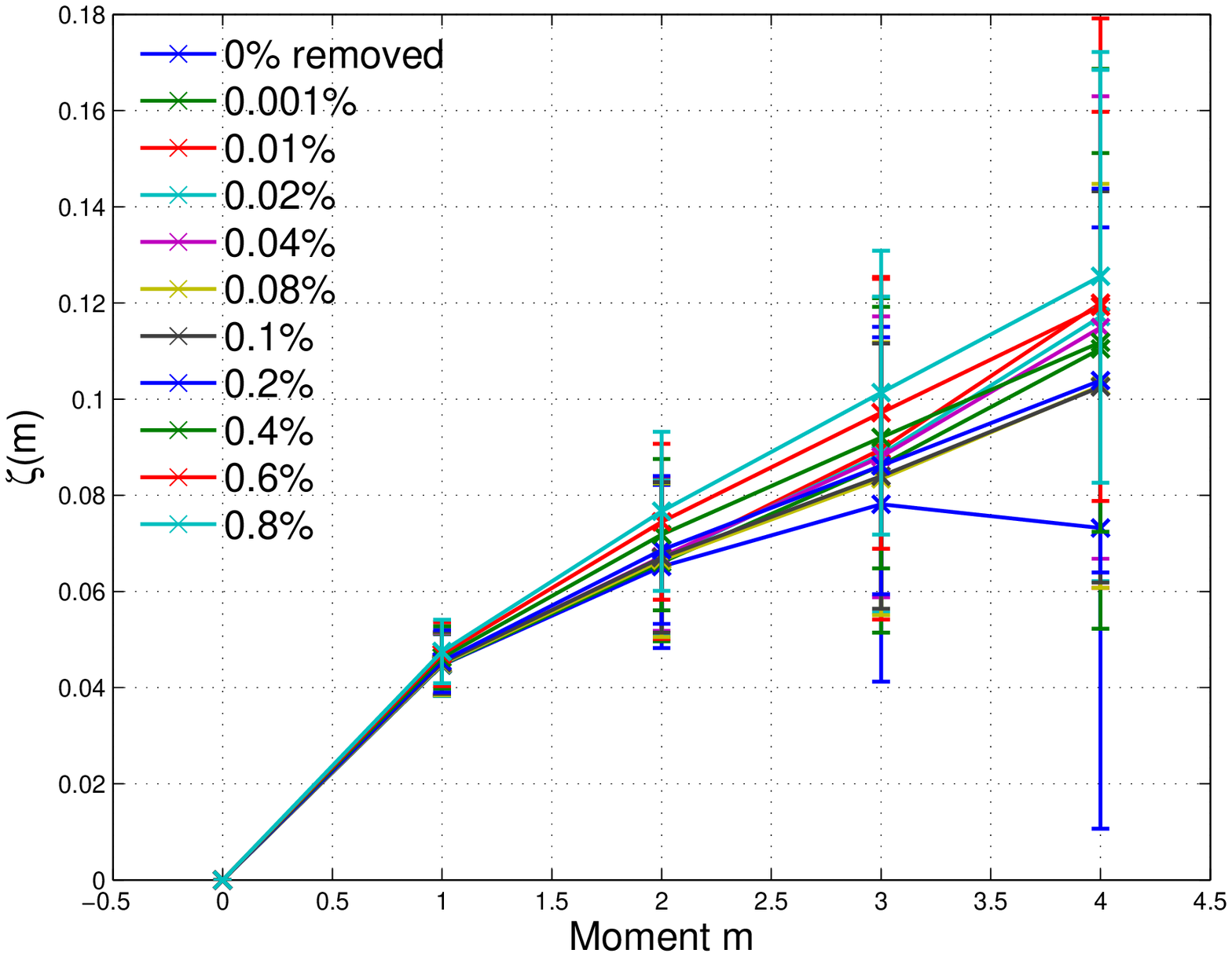}
\plotone{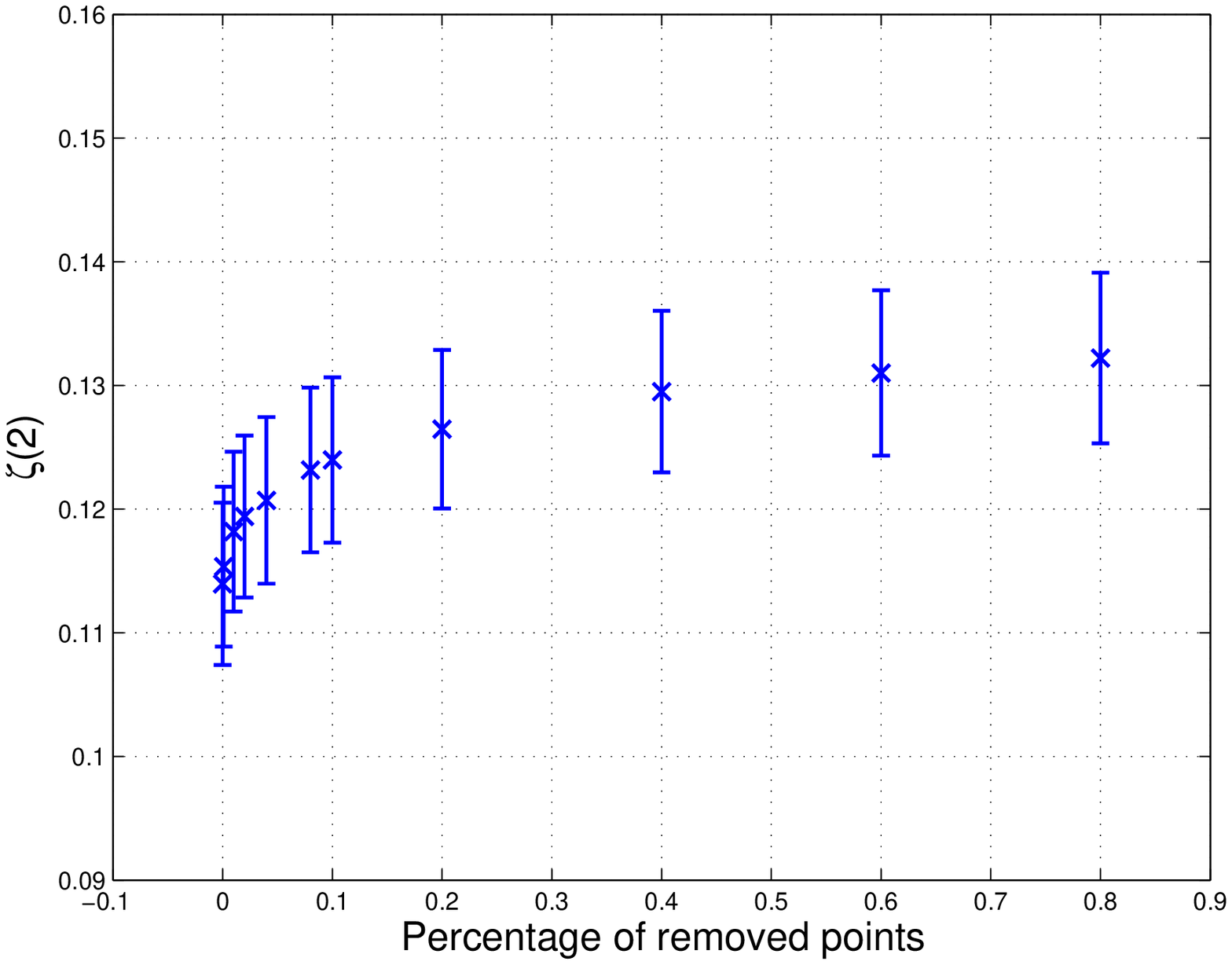}
\caption{Finite size effects on fractal and multifractal processes. The left-hand panels show the scaling exponents $\zeta(m)$ plotted as a function of moment $m=1$ to $6$ for different percentages of removed points. The right-hand panels show $\zeta(2)$ plotted against the percentage of removed points. The two top panels show the scaling for a Brownian walk, which has normally distributed steps; and the two bottom panels are the results for a multifractal p-model \citep{kiyani_07}.}
\label{Fig.10}
\end{figure}

We begin by considering a simple self-affine process where $S_{m}\sim\tau^{\zeta(m)}$, $\zeta(m)=Hm$. A fractal (self-affine) timeseries will in principle always give the same value of $H$ if computed from any region, or range of values, of the PDF of fluctuations (differences) sampled across a timescale $\tau$. We seek to choose the most statistically significant subset, and we do this by recomputing $H$ for different regions of the PDF; if the process is fractal, we expect to find the same $H$. To probe the full range of behaviour in the PDF, including any extended tails, we need to test for convergence to a single value of $H$ for a wide dynamic range of the PDF, for example $\sim20\sigma$. The largest values explored by the PDF of the data are the least well sampled statistically. It follows that if we successively remove these outliers, we should see, for a fractal timeseries, rapid convergence to a constant $H$ value. This is shown in Figure \ref{Fig.10} top panel for a Brownian walk, see also \citet{kiyani_07}. The scaling for a Brownian walk with normally distributed steps demonstrates the expected behaviour for a fractal process without heavy tails. On the other hand a multifractal process does not return a single constant value of $H$ as one changes the range of values over which $H$ is computed; this can be seen for the multifractal p-process \citep{meneveau_87} in the lower panel of Figure \ref{Fig.10}. A plot of the value of the exponent (here $\zeta(2)$) as we succesively remove outliers then can distinguish fractal and multifractal processes. For processes that are fractal, it also provides a more precise determination of the single exponent $H$ that characterizes the timeseries. The errors are obtained by combining the least squares error in the $\zeta(m)$ value fitted across the full range, with the standard deviation of the $\zeta(m)$ values fitted across runs of data points that have varying lengths, starting with a minimum length of about half the total fitting range length, centered on the middle of the full fitting range. This method is applied for all the exponent statistics throughout this paper.
\subsection{Fast quiet solar wind scaling}
\begin{figure}[H]
\figurenum{11}
\epsscale{0.4}
\plotone{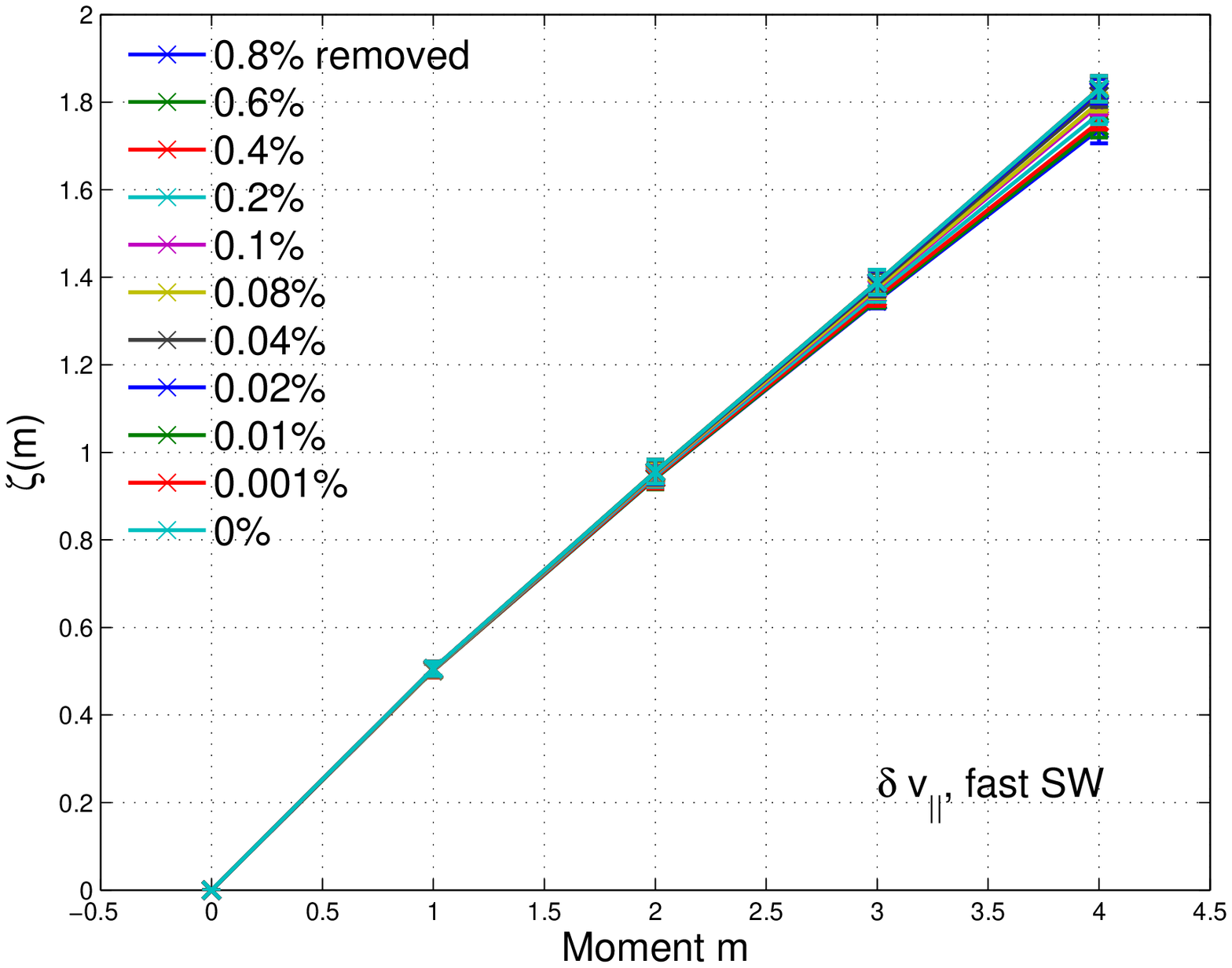}
\plotone{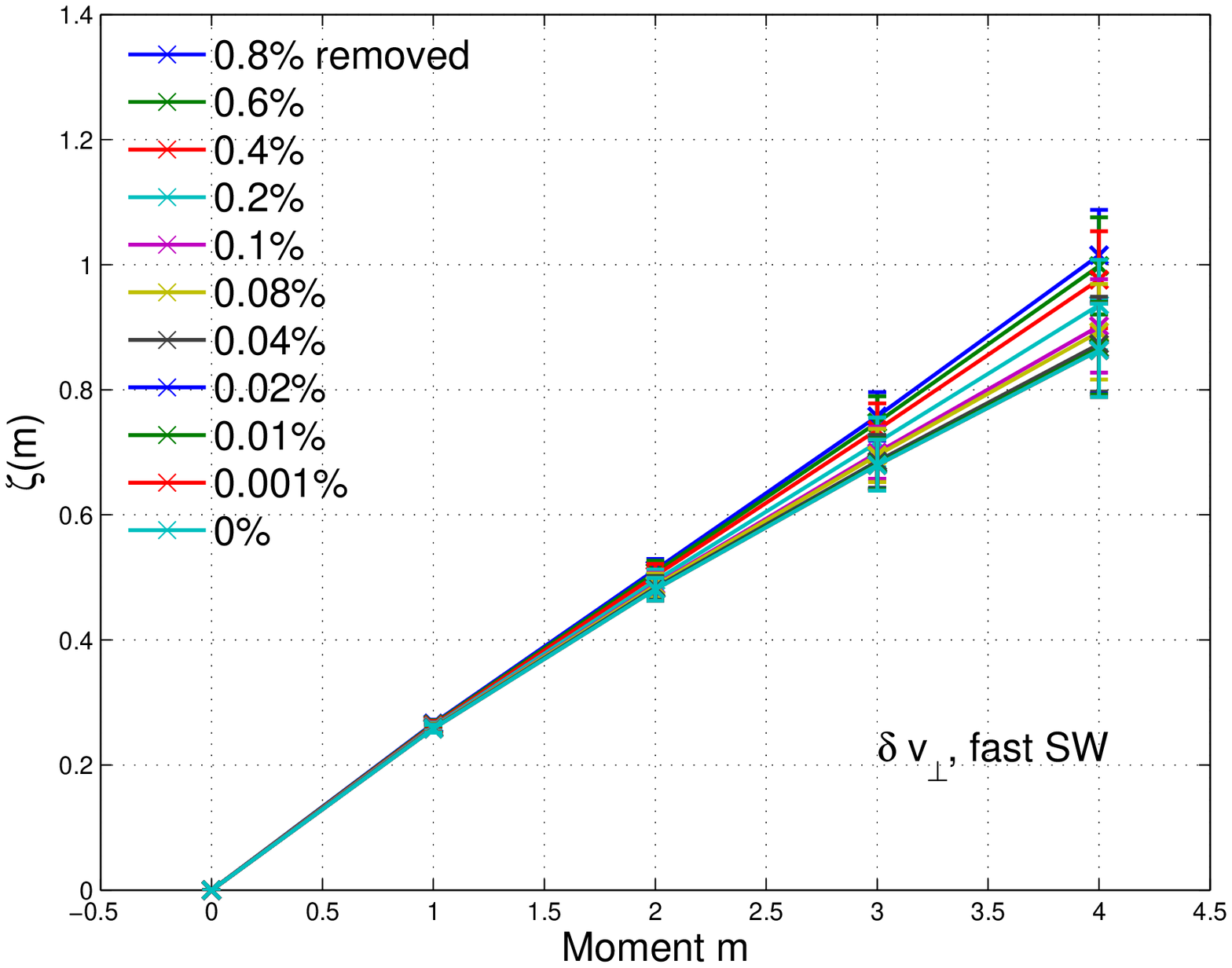}
\plotone{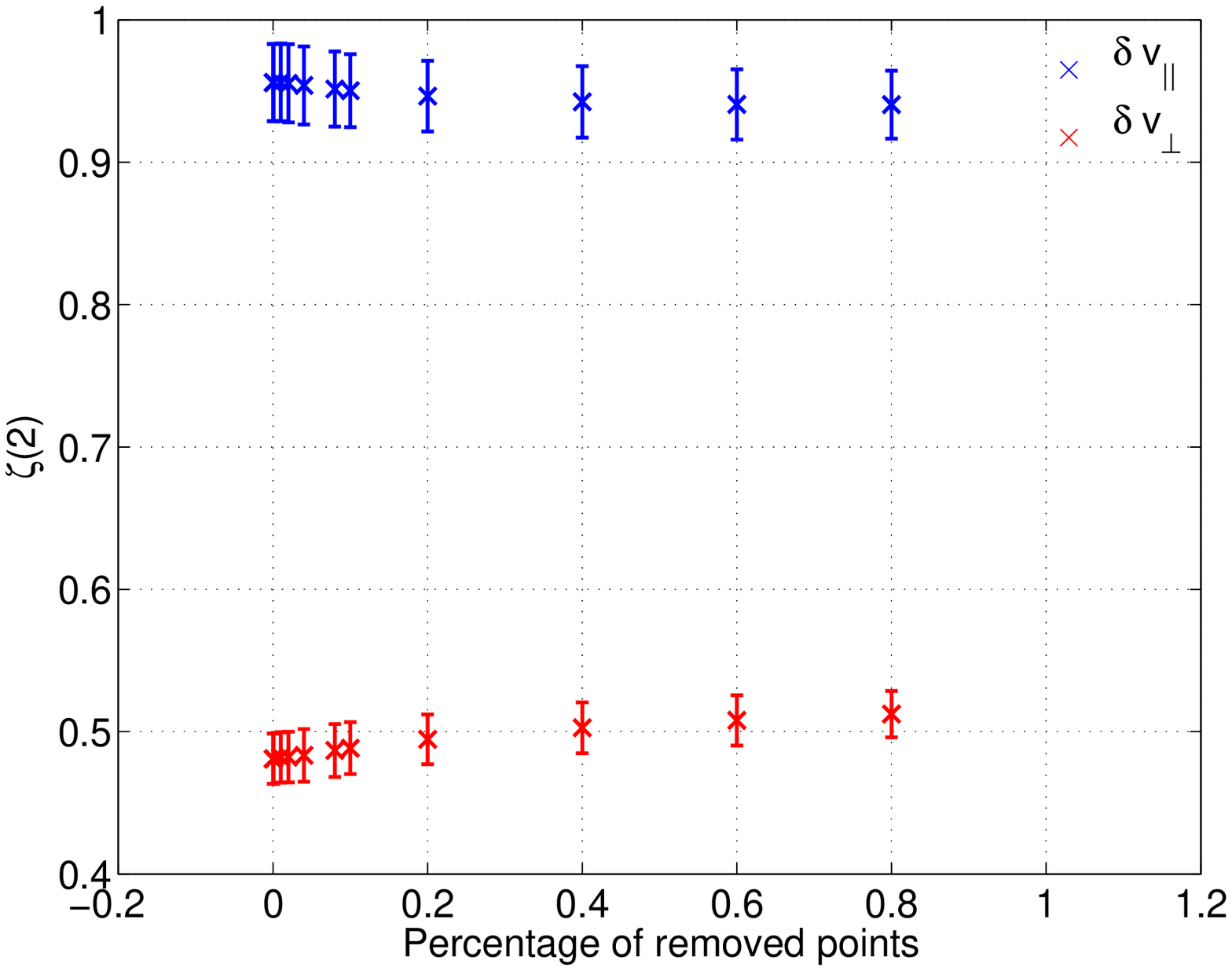}
\caption {Scaling properties of fluctuations $\delta v_{\parallel}$ and $\delta v_{\perp}$ in the ``$1/f$'' range, $\tau=320$ to $1002$ minutes, in the fast solar wind at solar minimum. The upper panels show the $\zeta(m)$ exponents plotted as a function of moment $m=1$ to $4$ for different percentages of removed points for $\delta v_{\parallel}$ (left) and $\delta v_{\perp}$ (right). The bottom panel shows $\zeta(2)$ plotted against the percentage of removed points for $\delta v_{\parallel}$ (blue, upper) and $\delta v_{\perp}$ (red, lower).}
\label{Fig.11}
\end{figure}

\begin{figure}[H]
\figurenum{12}
\epsscale{0.4}
\plotone{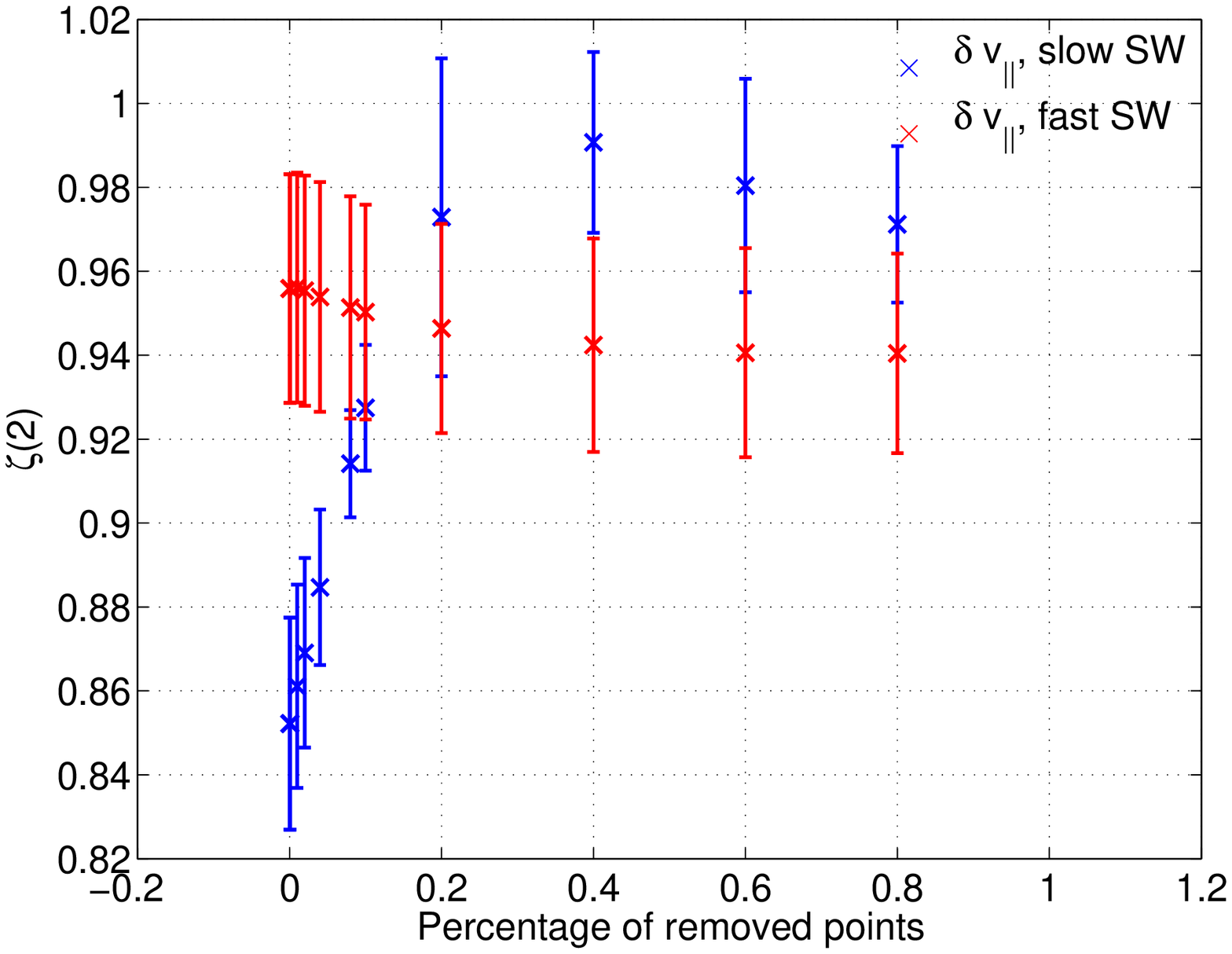}
\plotone{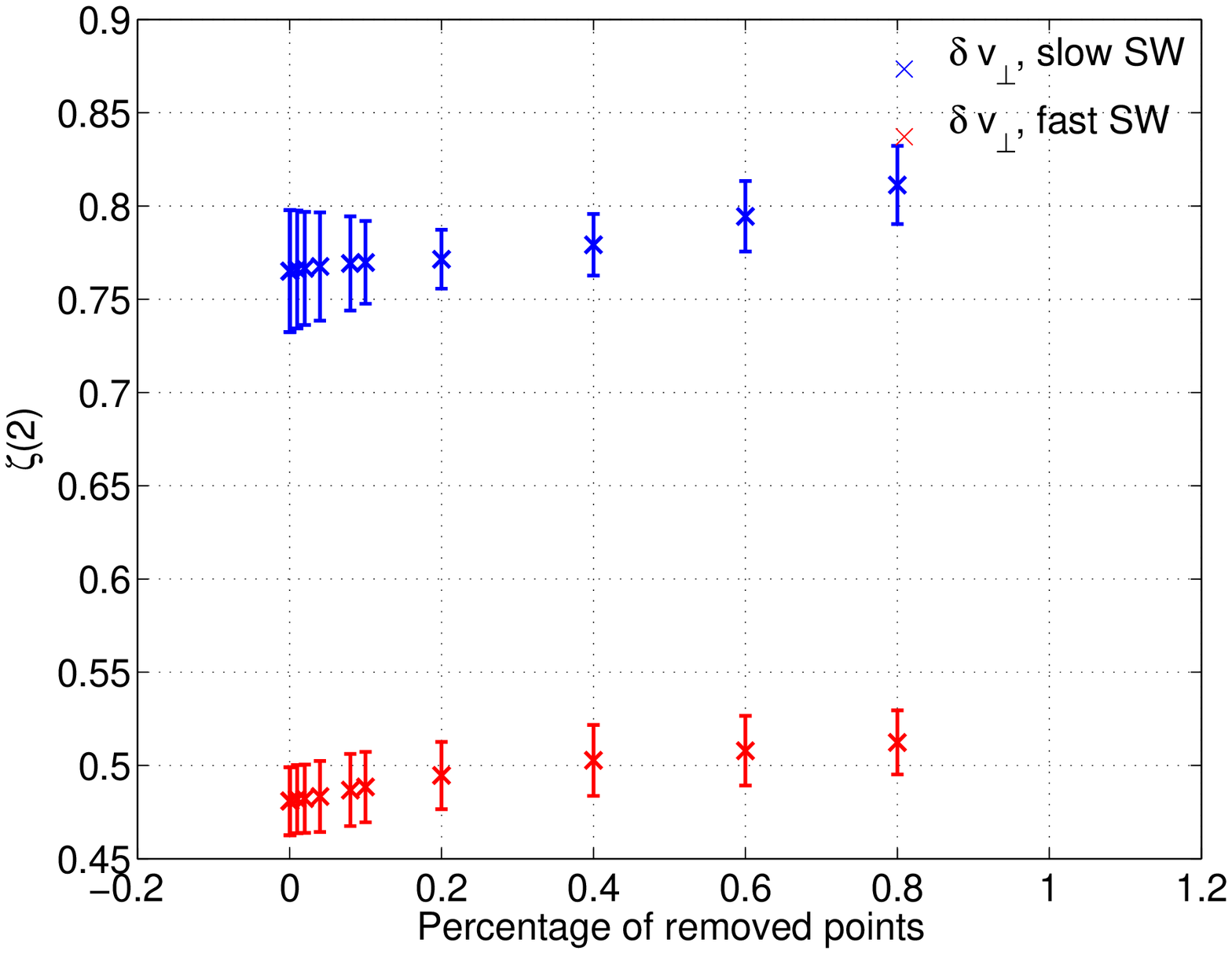}
\caption{Comparisons of $\zeta(2)$ at solar minimum in the ``$1/f$'' range as a function of the percentage of removed points for fast (red) and slow (blue) solar wind streams for $\delta v_{\parallel}$ (left) and $\delta v_{\perp}$ (right). The slow solar wind scaling appears to be more strongly multifractal.}
\label{Fig.12}
\end{figure}

We now quantify the scaling exponents of $\delta v_{\parallel}$ and $\delta v_{\perp}$ fluctuations in the fast solar wind at solar minimum. The corresponding GSFs are plotted in the top right panels of Figures \ref{Fig.8} and \ref{Fig.9}. We plot the exponents $\zeta(m)$, which are the gradients of the fitted power laws, from $\tau=320$ to $1002$ minutes in Figure \ref{Fig.11}. In the lower panels we show how the value of $\zeta(2)$ changes as outliers are successively removed. Comparing with Figure \ref{Fig.10}, we infer that $\delta v_{\parallel}$ is fractal within errors and $\delta v_{\perp}$ is only very weakly multifractal (almost monofractal). For the exponents, we obtain $\zeta_{\parallel}(2)$ close to $1$, suggestive of near Gaussian behaviour and (if the relation $\alpha=1+\zeta(2)$ holds) a PSD$\sim1/f^{2}$. In contrast, the exponent for perpendicular fluctuations $\zeta_{\perp}(2)$ is close to $0.5$, implying a PSD$\sim1/f^{3/2}$.
Figure \ref{Fig.12} compares the scaling exponents $\zeta(2)$ for $\delta v_{\parallel}$ and $\delta v_{\perp}$ in fast and slow solar wind streams at solar minimum. The corresponding GSFs are plotted in the right-hand pairs of panels in Figures \ref{Fig.8} and \ref{Fig.9}. Fluctuations of $\delta v_{\parallel}$ in the slow solar wind appear more strongly multifractal than in the fast wind. For $\delta v_{\perp}$ the slow solar wind displays a much higher exponent value for slow ($\zeta(2)\sim0.8$) than for fast ($\zeta(2)\sim0.5$) streams, reflecting the intrinsic differences between the fast and slow solar wind, and the coronal plasma conditions and magnetic field configuration at their origin.
To summarize the observed results: Analysis of the scaling exponents reveals fractal or weakly multi-fractal (very close to monofractal) scaling in the fluctuations of velocity components in the fast solar wind, with very different values for $\delta v_{\parallel}$ ($\zeta(2)\sim0.95$) and $\delta v_{\perp}$ ($\zeta(2)\sim0.5$) at solar minimum. In the slow solar wind at solar minimum, the scaling exponent $\zeta(2)$ of $\delta v_{\perp}$ nearly doubles from $\sim0.5$ to $\sim0.8$. In contrast, the scaling of $\delta v_{\parallel}$ remains quantitatively similar to its value in the fast wind (Figure \ref{Fig.12}), i. e. $\zeta(2)\sim0.95$ but has a less well-defined monofractal character. Finally, if we assume a regime in which the PSD $f^{-\alpha}$ scaling exponent $\alpha$ is related to $\zeta(2)$ by $\alpha=1+\zeta(2)$, then we obtain for the fast quiet solar wind: $\alpha\sim1$ for $\delta b_{\parallel,\perp}$ (as expected from Figure \ref{Fig.1}), $\alpha\sim2$ for $\delta v_{\parallel}$ and $\alpha\sim 1.5$ for $\delta v_{\perp}$.
\section{Conclusions}
We have examined the scaling of the parallel and perpendicular velocity and magnetic field fluctuations measured in the solar wind at $\sim1$ AU by ACE, which we have decomposed with respect to a locally averaged background magnetic field. Power spectra, GSFs and PDF collapse have been used to qualify and quantify the nature of the observed scaling in the low frequency ``$1/f$'' range. Slow and fast solar wind streams have been compared at both solar maximum in $2000$ and solar minimum in $2007$. The slow solar wind is found to be more multifractal and complex than the fast solar wind.\\
The magnetic field fluctuations display a flattening of the GSFs for $\tau\geq178$ minutes and a spectral index $\sim 1$, consistent with $\sim1/f$ behaviour found previously \citep{matthaeus_86, matthaeus_07}. In contrast, the velocity fluctuations show strong anisotropy, with scaling behaviour distinct from that of the $\mathbf{B}$ field and characterized by steepening of the GSFs in the ``$1/f$'' range (Figure \ref{Fig.7}) consistent with $\sim1/f^{\alpha},\;\alpha\neq1$.
\\For the fast quiet solar wind, $\delta v_{\parallel}$ and $\delta v_{\perp}$ have different scaling exponents: $\delta v_{\parallel}$ exhibits fractal scaling with $\zeta(2)\sim0.95\pm0.02$ whereas $\delta v_{\perp}$ is weakly multifractal with $\zeta(2)\sim0.49\pm0.03$ (Figure \ref{Fig.11}). The PDFs for these quantities also rescale relatively well. Also in the fast quiet solar wind, the PDF of $\delta v_{\parallel}$ is close to Gaussian, whereas $\delta b_{\parallel}$ is nearly symmetric and has stretched exponential tails, consistent with a multiplicative process. The rescaled PDFs for $\delta v_{\perp}$ and $\delta b_{\perp}$ in the fast solar wind can be fitted with the same distribution function, which is close to gamma or inverse Gumbel (see Figure \ref{Fig.5}). However their scaling exponents revealed by GSFs differ substantially (see Figure \ref{Fig.9}). This is consistent with a common coronal source for the fluctuations but a different spatiotemporal evolution out to $1$AU. The functional form of the PDF then constrains the mechanism that generates the fluctuations at the corona, gamma having points of contact with turbulence in confined plasmas \citep[see for example][and references therein]{graves_05,labit_07} and Gumbel, as an extremal process.\\
The breakpoint between the inertial range and ``$1/f$'' ranges differs between fast and slow solar wind streams and between periods of maximum and minimum solar activity. The inertial range extends to longer timescales in slow solar wind streams and at periods of maximum solar activity. The values of the inertial range scaling exponents remain unaffected by changes in the solar cycle (Figure \ref{Fig.8} and Figure \ref{Fig.9}), consistent with locally generated turbulence.\\
Our results clearly show very different behaviour between the magnetic and velocity fluctuations in the ``$1/f$'' range. The fractal nature of $\delta v_{\parallel}$ points to distinct physical processes in the corona, and to their mapping out into the solar wind. Further work would involve relating the fractal scaling observed at $\sim 1$AU with fractal stirring of magnetic footpoints in the corona. The different scaling observed in $\delta v_{\perp}$ points to different dynamics perpendicular to the background field (field line interactions?) with a possible common coronal origin the the $\delta v_{\perp}$ and $\delta b_{\perp}$ fluctuations. in fast quiet solar wind.
\section{Acknowledgments}
RN acknowledges the STFC and UKAEA Culham for financial support and R. P. Lepping and the ACE team for data provision.

\clearpage
\end{document}